\documentclass[journal,twoside,10pt]{IEEEtran}
\usepackage{ifpdf}
\usepackage{cite}
\ifCLASSINFOpdf
\usepackage[pdftex]{graphicx}
\DeclareGraphicsExtensions{.pdf,.jpeg,.png}
\else
\usepackage[dvips]{graphicx}
\DeclareGraphicsExtensions{.eps}
\fi
\graphicspath{{./Figures/}}
\usepackage{graphicx}
\ifCLASSOPTIONcompsoc
\usepackage[caption=false, font=scriptsize, labelfont=sf, textfont=sf]{subfig}
\else
\usepackage[caption=false, font=footnotesize]{subfig}
\fi
\usepackage{floatrow}
\usepackage{algorithm,algorithmic} 
\usepackage{caption}
\usepackage{bbm}
\usepackage{balance}
\usepackage{epstopdf}
\usepackage{amsmath,amsfonts,bm,amssymb}
\usepackage{color}
\usepackage{array}
\usepackage{mathtools} 
\usepackage{filecontents} 
\usepackage{multirow} 
\usepackage{booktabs} 
\usepackage{diagbox}
\usepackage{makecell}
\usepackage{url}
\usepackage{cite}
\usepackage{cleveref}
\usepackage{tabularx} 

\newcommand{\bs}[1]{\boldsymbol{#1}}
\newcommand{\mc}[1]{\mathcal{#1}}

\newcommand{\mb}[1]{\mathbf{#1}}
\newcommand{\mr}[1]{\mathrm{#1}}
\newcommand{\ms}[1]{\mathsf{#1}}
\newcommand{\tr}{\mathrm{Tr}}
\newcommand{\e}{\mathrm{e}}

\newcommand{\lr}[1]{\langle #1 \rangle}
\newcommand{\blr}[1]{\big\langle #1 \big\rangle}
\newcommand{\bblr}[1]{\bigg\langle #1 \bigg\rangle}
\newcommand{\Blr}[1]{\Big\langle #1 \Big\rangle}

\newcommand{\low}{\mr{low}}
\newcommand{\up}{\mr{up}}
\definecolor{green}{rgb}{0.0, 0.5, 0.0}
\definecolor{grey}{rgb}{0.52, 0.52, 0.51}
\usepackage{setspace}
\usepackage{lipsum}

\usepackage{empheq}
\begin{document}
%
\newcommand{\C}{\mathbb{C}}
\newcommand{\R}{\mathcal{R}}
\newcommand{\RR}{\mathbb{R}}
\newcommand{\A}{\mathbf{A}}
\newcommand{\n}{\mathbf{n}}
\newcommand{\I}{\mathcal{I}}
\newcommand{\Ib}{\mathbf{I}}
\newcommand{\X}{\mathbf{X}}
\newcommand{\Y}{\mathbf{Y}}
\newcommand{\Z}{\bm{Z}}
\newcommand{\x}{\mathbf{x}}
\newcommand{\si}{\mathbf{s}}
\newcommand{\sd}{\Sigma\Delta}
\newcommand{\st}{\rm st}
\newcommand{\nd}{\rm nd}
%
%
%
%
\newcommand{\h}{\mathbf{h}}
\newcommand{\g}{\mathbf{g}}
\newcommand{\ve}{\mathrm{v}}
\newcommand{\V}{\mathbf{V}}
\newcommand{\vv}{\mathbf{v}}
\newcommand{\w}{\mathbf{w}}
%
\newcommand{\alp}{\boldsymbol\alpha}
\newcommand{\bta}{\boldsymbol\beta}
\newcommand{\gma}{\boldsymbol\gamma}
\newcommand{\taub}{\boldsymbol\tau}
\newcommand{\varthetab}{\boldsymbol\vartheta}
\newcommand{\varphib}{\boldsymbol\varphi}
\newcommand{\sinr}{\mathtt{SINR}}

\newcommand{\sn}[1]{{\color{blue} \bf{{{{SN --- #1}}}}}}
\setlength{\abovedisplayskip}{2.0pt}
\setlength{\belowdisplayskip}{2.0pt}
%
\newcommand{\note}[1]{{\fontfamily{cmss}\selectfont\color{red}\noindent\text{\textbf{Note}: }{#1}}}
\title{MIMO Detection with Spatial Sigma-Delta ADCs: A Variational Bayesian Approach}
\author{Toan-Van Nguyen, \emph{Member, IEEE}, Sajjad Nassirpour, \emph{Member, IEEE}, Italo Atzeni, \emph{Senior Member, IEEE}, Antti T\"olli, \emph{Senior Member, IEEE}, A. Lee Swindlehurst, \emph{Fellow, IEEE} and Duy H. N. Nguyen, \emph{Senior Member, IEEE}
\thanks{Toan-Van Nguyen, Sajjad Nassirpour, and Duy H. N. Nguyen are with the Department of Electrical and Computer Engineering, San Diego State University, San Diego, CA 92182 USA (emails:\{tnguyen58, snassirpour, duy.nguyen\}@sdsu.edu).}
\thanks{Italo Atzeni and Antti T\"olli are with the Centre for Wireless Communications, University of Oulu, Finland (emails:\{italo.atzeni, antti.tolli\}@oulu.fi).}
\thanks{A. Lee Swindlehurst is with the Department of Electrical Engineering and Computer Science, University of California, Irvine, CA, USA (email: swindle@uci.edu).}
\thanks{Part of this paper has been published at the 2024 Asilomar Conference on Signals, Systems, and Computers, Pacific Grove, CA, USA, Oct. 2024 \cite{van2024Asilomar}.}}
\maketitle
\IEEEpeerreviewmaketitle
\begin{abstract}
The spatial Sigma-Delta ($\sd$) architecture can be leveraged to reduce the quantization noise and enhance the effective resolution of few-bit analog-to-digital converters (ADCs) at certain spatial frequencies of interest. Utilizing the variational Bayesian (VB) inference framework, this paper develops novel data detection algorithms tailored for massive multiple-input multiple-output (MIMO) systems with few-bit $\Sigma\Delta$ ADCs and angular channel models, where uplink signals are confined to a specific angular sector. We start by modeling the corresponding Bayesian networks for the $1^{\st}$- and $2^{\nd}$-order $\Sigma\Delta$ receivers. Next, we propose an iterative algorithm, referred to as Sigma-Delta variational Bayes (SD-VB), for MIMO detection, offering low-complexity updates through closed-form expressions of the variational densities of the latent variables. {We also study the impact of mutual coupling on the performance of the proposed SD-VB algorithms when the antenna spacing is reduced.} Simulation results show that the proposed $2^{\nd}$-order SD-VB algorithm delivers the best symbol error rate (SER) performance while maintaining the same computational complexity as in unquantized systems, matched-filtering VB with conventional quantization, and linear minimum mean-squared error (LMMSE) methods. Moreover, the $1^{\st}$- and $2^{\nd}$-order SD-VB algorithms achieve their lowest SER at an antenna separation of one-fourth wavelength for a fixed number of antenna elements. The effects of  {mutual coupling,} the steering angle of the $\sd$ architecture, the number of ADC resolution bits, and the number of antennas and users are also extensively analyzed.
\end{abstract}
\section{Introduction} \label{sec_introduction}
Beyond-5G wireless systems will require substantial bandwidth in both the millimeter (mmWave) and terahertz (THz) bands to deliver high data throughput \cite{rajatheva2020white}. Signals at these frequencies, however, are hindered by low penetration capabilities and high propagation loss, which restrict their practical communication range \cite{nguyen2017hybrid}.
Massive multiple-input multiple-output (MIMO) arrays have been used to compensate for the propagation loss while simultaneously achieving high capacity through spatial multiplexing \cite{buzzi2018energy}. However, exploiting the full benefits of beamforming and multiplexing in massive MIMO can be challenging due to the need for dedicated high-resolution analog-to-digital converters (ADCs)/digital-to-analog converters (DACs) for each antenna element \cite{sohrabi2018one,jedda2018}. This results in high hardware complexity and increased power consumption, especially with larger bandwidths and sampling rates \cite{fesl2023mean}.
To address these concerns, the use of low-resolution ADCs with, e.g., $1$--$3$ bits of precision, has emerged as an energy-efficient and low-complexity solution for massive MIMO systems \cite{roth2018comparison,safa2024data,nguyen2023variational}.

Low-resolution ADCs provide low-complexity and low-power designs for massive MIMO systems, but present challenges due to their non-linear nature \cite{li2017channel}. The impact of low-resolution ADCs on MIMO systems has been explored, including analyses of the resulting capacity limits and achievable rates \cite{zhang2016spectral,fan2015uplink,saxena2017analysis,li2017channel,bazrafkan2020asymptotic}. These studies show that low-resolution ADCs degrade system performance metrics such as achievable rate and symbol error rate (SER), especially at medium-to-high signal-to-noise ratios (SNRs), but this negative impact can be mitigated by increasing the number of antennas.
This observation indicates that massive MIMO can be effectively operated with few-bit ADCs. However, attaining a good tradeoff between the system performance and the number of quantization bits requires advanced signal processing algorithms and architectural innovations tailored for the specific characteristics of the quantized signals \cite{nguyen2021svm,nguyen2023variational}.

Sigma-Delta ($\sd$) modulation is a well-established method for time-domain signal processing that encodes analog signals into low-bit-depth digital signals at a very high sampling rate. By combining oversampling and noise shaping, $\sd$ modulation can achieve high resolution despite using low-bit quantizers \cite{de2010sigma,rao2021massive}. In temporal $\sd$, oversampling involves sampling a signal at a rate higher than the Nyquist rate. 
A negative feedback loop is employed to quantize the difference between the input signal and the quantized output to effectively shape the quantization noise to higher frequencies \cite{de2010sigma}. A low-pass filter is then used to remove the high-frequency quantization noise while preserving the desired low-frequency signals 
\cite{aziz1996overview}. The $1^{\st}$-order temporal $\sd$ approach is widely favored in data-driven applications that demand high precision and low noise, such as audio recording, medical imaging and diagnostics, industrial sensor systems, as well as mobile devices used for signal processing \cite{de2010sigma}.
In massive MIMO, spatial $\sd$ quantization has been used to enhance channel estimation and spectral efficiency \cite{rao2021massive,pirzadeh2020spectral}.
In spatial $\sd$, the difference between the input and the quantized output represents the quantization noise at each antenna, which is fed back and compared with the input to an adjacent antenna. This feedback enables more aggressive noise shaping, where the spatial spectrum of the quantization noise is pushed away from the spatial frequencies of interest \cite{sankar2022channel}.

\subsection{Related Works}
Communications in the mmWave and THz bands typically require large antenna arrays at the base stations (BS). Here, the implementation of one-bit ADCs for every antenna element can reduce radio-frequency complexity, cost, and power consumption \cite{atzeni2023doubly,jeon2018one}.  
In \cite{jeon2019robust}, a likelihood function learning method for data detection was proposed with one-bit ADCs using an approach based on reinforcement learning. In \cite{nguyen2021svm,nguyen2020svm}, a support vector machine (SVM)-based channel estimation and data detection method was exploited for one-bit massive MIMO systems, yielding a data detection performance close to that of maximum-likelihood detection. Data detection based on supervised learning was proposed for few-bit MIMO systems in \cite{jeon2018supervised}, whereas an approach based on deep neural networks was developed in \cite{khobahi2021lord}. These learning-based methods necessitate a significant training load, and the neural network must be retrained with new channel realizations \cite{khobahi2021lord}, thereby increasing the computational complexity. To reduce computational complexity, zero-forcing-based data detection was studied in uplink massive MIMO systems with few-bit ADCs in \cite{azizzadeh2019ber}.

{Generalized approximate message passing (GAMP)-based algorithms, such as generalized orthogonal approximate message passing (GOAMP) \cite{liu2023achievable} and generalized memory approximate message passing (GMAMP) \cite{tian2022generalized}, have also been proposed for generalized linear models with few-bit ADCs. These algorithms can be applied to any discrete input signals and achieve Bayesian optimality when the transformation matrix is high-dimensional and is unitarily invariant}. A state evolution framework was used in these works to analyze the performance of GAMP-based algorithms in the large-system limit. {However, for mmWave channel models with non-unitarily invariant channel structures and highly non-linear, non-separable spatial $\sd$ quantization, there is no guarantee that GOAMP and GMAMP will converge or achieve Bayesian optimality, motivating the need for alternative detection approaches.} 
The use of analog comparator networks prior to 1-bit quantization was introduced in \cite{fernandes2022multiuser} to effectively create virtual channels whose outputs can enhance the performance of MIMO receivers. To further reduce the effect of non-linear quantization, the authors of \cite{chen2024low} studied the integration of pre-ADC comparators and deep learning detection in orbital angular momentum systems with 1-bit spatial oversampling.

To efficiently balance the performance and complexity for systems with few-bit ADCs, variational Bayesian (VB) inference, inspired by machine learning, was proposed for data detection and channel estimation in \cite{nguyen2023variational}. The goal of VB is to find an approximation for the true posterior distribution of latent variables given the observed data \cite{nguyen2022variational}. In \cite{zhu2019grid}, the VB method was applied for gridless channel estimation in antenna array systems with low-resolution ADCs. 


Recently, spatial $\sd$ ADCs in massive MIMO systems have attracted growing interest in terms of channel estimation \cite{rao2021massive,sankar2022channel}, performance analysis \cite{rao2021massive,pirzadeh2020spectral,palguna2016millimeter}, and precoding design \cite{shao2019one}. In \cite{pirzadeh2020spectral}, the spectral efficiency of $1$-bit $\sd$ massive MIMO was analyzed, revealing that one-bit $\sd$ scales down the quantization noise power proportionally to the square of the spatial oversampling rate. Spatial few-bit $\sd$ ADCs have also been used in massive MIMO to shape the quantization noise away from users in certain angular sectors for improved channel estimation \cite{rao2021massive}. In \cite{palguna2016millimeter}, a parallel $\sd$ ADC was designed for mmWave receivers to achieve high-resolution outputs; however, the bit-error-rate performance of this design with $3$-bit quantization was not significantly improved at high SNRs.
$1^{\st}$- and $2^{\nd}$-order $\sd$ quantization was employed for wide-band systems in \cite{corey2016spatial} without considering the angle steering mechanism.
A reconfigurable intelligent surface controlled by a single-antenna base station equipped with $1^{\st}$-order $\sd$ modulation was investigated in \cite{keung2024transmitting}. Mutual coupling effects for phase quantization were considered for dense $\sd$ phased arrays in \cite{krieger2013dense}.  {A VB algorithm for optimal detection in MIMO systems using spatial $\sd$ ADCs was recently developed in \cite{van2024Asilomar}. This approach leverages the noise-shaping properties of the $1^{\rm st}$-order $\sd$ architecture to improve detection performance in angularly constrained scenarios. However, the study remains limited to $1^{\rm st}$-order $\sd$ ADCs, which do not fully capitalize on the potential performance gains offered by higher-order $\sd$ architectures.}

\subsection{Motivations and Contributions}
A common approach in the aforementioned works on $\sd$ ADCs is to exploit the Bussgang decomposition, which reformulates the nonlinear quantization as a linear function, followed for example by standard linear data detection and channel estimation methods \cite{li2017channel,shao2019one,rao2021massive}. 
 {However, the Bussgang decomposition relies on the assumption of Gaussian signals, which results in a diagonal Bussgang gain matrix. At high SNR with specular channels \cite{andrews2016modeling} and relatively few users, the Gaussian assumption does not hold and the Bussgang gain matrix becomes non-diagonal, leading to correlated distortion terms across users or spatial dimensions \cite{demir2020bussgang,alkhateeb2014channel,rao2021massive}.}
 {For example, this effect is manifested when the nonlinear quantization generates harmonics for sparse specular inputs, as observed in \cite{jin2020one}. 
}

The VB method has been proposed as an alternative to the Bussgang approach to effectively address the non-linear quantization operation. It is considered superior to linear methods, bilinear GAMP \cite{thoota2021variational}, and other approaches based on deep learning \cite{nguyen2022deep}. However, these studies are limited to conventional few-bit ADCs.  {More importantly, achieving optimal MIMO detection with spatial $\sd$ ADCs under the maximum-a-posteriori (MAP) probability criterion may not be feasible due to the complex noise-shaping effects that disrupt the posterior distribution. Building on our initial exploration of MIMO detection with $1^{\st}$-order $\sd$ ADCs
in \cite{van2024Asilomar}, this paper develops a comprehensive VB framework for MIMO detection that incorporates both $1^{\st}$- and $2^{\nd}$-order $\sd$ architectures. This framework addresses key limitations of prior VB studies, such as their focus on conventional ADCs and lack of support for higher-order noise shaping.}
 {Moreover, we consider a narrowband channel model typical of mmWave systems, where the signal bandwidth is smaller than the coherence bandwidth. This model is relevant to applications such as short-range indoor communications \cite{heath2016overview}, mmWave MIMO integrated sensing and communication systems, and low-mobility IoT networks \cite{kang2023deep}, where LoS paths or narrow angular sectors dominate over frequency selectivity. The VB inference is used to develop novel detection algorithms tailored for such scenarios.}

The contributions of the paper are summarized as follows: 
\begin{itemize}
 \item  {We develop $1^{\rm st}$- and $2^{\rm nd}$-order \underline{S}igma-\underline{D}elta \underline{V}ariational \underline{B}ayes (SD-VB) detection algorithms for massive MIMO systems with few-bit $\Sigma\Delta$ quantizers. To support this, we employ a Bayesian network to model the $\Sigma\Delta$ receiver, capturing insightful dependencies between the observed and latent variables for effective inference.}
%
    \item We develop the $1^{\st}$- and $2^{\nd}$-order SD-VB algorithms for the $1^{\st}$- and $2^{\nd}$-order $\sd$ ADC architectures with efficient updates based on the analysis of the variational densities.  {Considering mutual coupling (MC) in $\sd$ architectures, we also present an LMMSE detector for MIMO detection with $1^{\st}$-order $\sd$ quantization \cite{pirzadeh2020spectral}.}
     The results show that the $1^{\st}$- and $2^{\nd}$-order SD-VB algorithms have the same computational complexity as the matched-filter quantized VB (MF-QVB) algorithm \cite{nguyen2023variational}, and significantly lower complexity than the LMMSE detector. 
    \item We demonstrate through simulation results that the $2^{\nd}$-order SD-VB algorithm achieves the lowest SER compared with state-of-the-art detection algorithms such as MF-QVB and LMMSE for certain angular sectors.  {We also examine the impact of mutual coupling on the detection performance of all considered algorithms.}
    Moreover, the $1^{\st}$-order SD-VB algorithm is more efficient than $2^{\nd}$-order SD-VB for $1$-bit ADCs.
\end{itemize}

The rest of the paper is organized as follows. Section~\ref{sec:Sys} introduces the system model as well as the model for few-bit $\sd$ ADCs. The optimal solution for the LMMSE detector based on the linearized model is derived in Section~\ref{sec:LMMSE}. Section~\ref{sec:1stSD} analyzes and discusses the VB framework for detection in a $1^{\st}$-order $\sd$ MIMO system. Then, Section~\ref{sec:2ndSD} extends the VB framework to the $2^{\nd}$-order $\sd$ case. Numerical results are provided in Section~\ref{sec:Numerical} and Section~\ref{sec:Conclusions} concludes the paper. 

\underline{\textit {Notation}}: Boldface lowercase and boldface uppercase variables denote vectors and matrices, respectively. The $i$-th element of a vector $\x$ is represented by $[\x]_i$ and the $(i,j)$-th element of a matrix $\X$ is denoted by $[\X]_{i,j}$. The symbols $\C$ and $\RR$ stand for the sets of complex and real numbers, respectively. The $L_2$-norm and the absolute value are indicated by $\|\cdot\|$ and $|\cdot|$, respectively. Real and imaginary parts are denoted by $\mathfrak{R}\{\cdot\}$ and $\mathfrak{I}\{\cdot\}$, respectively, with $j=\sqrt{-1}$. 
The identity matrix is denoted by $\mb{I}$, the trace operator by $\tr$($\cdot$), and the expectation operator by $\mathbb{E}\{\cdot\}$. The transpose, complex conjugate, and complex conjugate transpose operators are denoted by $(\cdot)^T$, $(\cdot)^*$, and $(\cdot)^H$, respectively. The symbols $\propto$ and $\sim$ stand for ``is proportional to'' and ``distributed according to'', respectively. The prior distribution function and variational distribution function of a random variable $x$ are denoted as $p(x)$ and $q(x)$, respectively. The notation $\mc{CN}(\bs{\mu},\bs{\Sigma})$ stands for a Gaussian random vector with mean $\bs{\mu}$ and covariance $\bs{\Sigma}$, whereas $\mc{CN}(\mb{x};\bs{\mu},\bs{\Sigma}) = \frac{1}{\pi^K|\bs{\Sigma}|}\,\mr{exp}\big(-(\mb{x}-\bs{\mu})^H\bs{\Sigma}^{-1}(\mb{x}-\bs{\mu})\big)$ denotes the probability density function of $\mb{x}\sim \mc{CN}(\bs{\mu},\bs{\Sigma})$. The bracket $\lr{.}$ denotes the expectation with respect to all latent variables except the one under consideration. The cosine and sine integrals are denoted as $\mathrm{Ci}(x)\triangleq \gamma + \mathrm{log}(x)+\int_{0}^{x}{\frac{\mathrm{cos}(t) - 1}{t}dt}$, and $\mathrm{Si}(x)\triangleq\int_{0}^{x}{\frac{\mathrm{sin}(t)}{t}dt}$, where $\gamma$ is the Euler-Mascheroni constant.
\vspace{-0.5em}
\section{System Model and Prior Art} \label{sec:Sys}
\subsection{Channel Model}
\begin{figure}[!t]
	\centering
\includegraphics[width=0.9\linewidth]{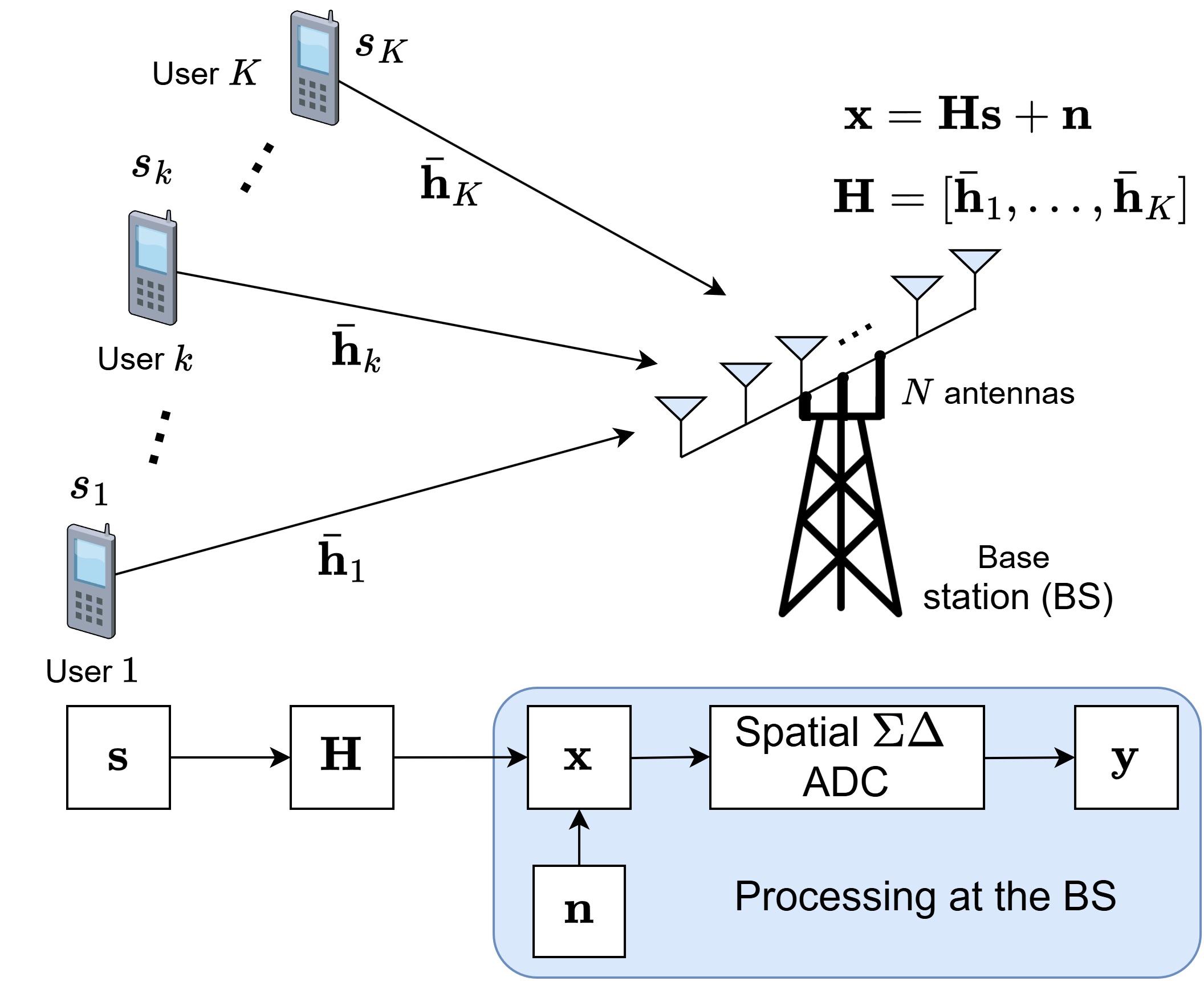}
	\caption{{Diagram of the considered system model where $K$ single-antenna users are transmitting to an $N$-antenna base station employing $\Sigma\Delta$ quantization.}}
	\label{SD_VB} 
\end{figure}

\begin{table}[ht]
\centering
\caption{List of main notations.}
\label{tab:parameters}
\begin{tabular}{p{1cm} p{6cm}}
\toprule
\textbf{Symbol} & \textbf{Description} \\
\midrule
$K$ & Number of users \\
$N$ & Number of antennas at the BS \\
$\mathbf{x}$ & Received signal vector before quantization \\
$\mathbf{H}$ & Uplink channel matrix, $\mathbb{C}^{N \times K}$ \\
$\mathbf{s}$ & Uplink data symbol vector, $\mathbb{C}^{K \times 1}$ \\
$\mathbf{n}$ & Additive Gaussian noise, $\mathcal{CN}(\mathbf{0}, N_0 \mathbf{I}_N)$ \\
$N_0$ & Noise variance \\
$\mathcal{S}$ & Discrete constellation set (e.g., QAM, PSK) \\
$L$ & Number of propagation paths per user \\
$\mathbf{A}_k$ & Array response matrix for user $k$, $\mathbb{C}^{N \times L}$ \\
$\mathbf{g}_k$ & Small-scale fading vector, $\mathcal{CN}(\mathbf{0}, \mathbf{I}_L)$ \\
$\beta_k$ & Geometric attenuation and slow fading \\
$\theta_{k\ell}$ & AoA of the $\ell$-th path from user $k$ \\
$\omega_{k\ell}$ & Spatial frequency, $2\pi \frac{d}{\lambda} \sin \theta_{k\ell}$ \\
$d$ & Inter-antenna spacing of the ULA \\
$\lambda$ & Wavelength \\
$\mathcal{S}_{\theta_0}$ & Angular sector, $[\theta_0 - \Theta/2, \theta_0 + \Theta/2]$ \\
$\theta_0$ & Center angle of the angular sector \\
$\Theta$ & Azimuth angular spread \\
$\mathbf{T}$ & Mutual coupling matrix, $\mathbb{C}^{N \times N}$ \\
$\mathbf{R}$ & Covariance matrix of colored noise \\
$b$ & Number of bits in ADC quantization \\
$\mathcal{Q}_b(\cdot)$ & $b$-bit quantization function \\
$\Lambda$ & Quantization step size \\
$d_m$ & Quantization threshold, $m \in \{1, \ldots, 2^b-1\}$ \\
$r_i$ & Pre-quantized signal at antenna $i$ \\
$\phi$ & Phase shift between adjacent antennas \\
$y_i$ & Quantized observation at antenna $i$ \\
$\mathbf{q}$ & Effective quantization noise vector \\
$\mathbf{U}$ & Lower triangular matrix for $\Sigma\Delta$ linearization \\
$\tau_{q_i}$ & Variance of quantization noise at antenna $i$ \\
$\mathbf{W}$ & LMMSE receiver matrix \\
$\mathbf{\Sigma}_{\mathbf{s}}$ & Covariance matrix of $\mathbf{s}$ \\
$\mathbf{\Sigma}_{\mathbf{q}}$ & Covariance matrix of $\mathbf{q}$ \\
$\gamma$ & Precision parameter, $1/N_0^{\text{post}}$ \\
$\langle r_i \rangle$ & Variational mean of $r_i$ \\
$\tau_{r_i}$ & Variational variance of $r_i$ \\
\bottomrule
\end{tabular} \vspace{-2em}
\end{table}

 {We consider an uplink massive MIMO system with single-carrier spatial-division multiple access as shown in Fig. \ref{SD_VB}, where $K$ single-antenna users transmit their signals simultaneously to a BS equipped with a uniform linear array (ULA) of $N$ antennas.}
The received signal $\x$ at the BS before quantization can be expressed as
\begin{align} \label{eq:x}
    \x = \mb{Hs} + \n,
\end{align}
where $\mb{H} = [\h_1,\ldots,\h_N]^T \in \mathbb{C}^{N \times K}$ is the uplink channel matrix, with $\h_i \in \mathbb{C}^{K\times 1}$ denoting the channel vector from the $K$ users to the $i$-th antenna at the BS, $\mb{s} = [s_1,\ldots,s_K]^T \in \mathbb{C}^{K\times 1}$ is the uplink data symbol vector, and $\n \sim \mc{CN}(\mb{0}, N_0 \mb{I}_N)$ represents additive Gaussian noise at the BS with variance $N_0$. We assume that the channel $\mb{H}$ is known at the BS. The symbol $s_k$ transmitted from user $k$ is drawn from a complex-valued discrete constellation $\mc{S}$, e.g., quadrature amplitude modulation (QAM) or phase-shift keying (PSK), with $\mathbb{E}[s_k] = 0$ and normalized such that $\mathbb{E}[|s_k|^2] = 1$. 
The prior distribution of $s_k$ can be expressed as
\begin{align}
    p(s_k) = \sum_{a \in \mc{S}}p_a\delta(s_k - a),
\end{align}
where $p_a$ is the known prior probability of the constellation point $a$ and $\delta(s_k-a)$ indicates the point mass function at $a$. 

We assume a geometric channel model typical of mmWave communications systems in which the channel for each user is composed of a linear combination of $L$ propagation paths \cite{alkhateeb2014channel,nguyen2017hybrid}, i.e.,
\begin{align} \label{eq:h_k}
    \bar{\mb{h}}_k = \sqrt{{\beta_k}/{L}} \mb{A}_k \mb{g}_k,
\end{align}
where $\bar{\mb{h}}_k$ is the $k$th column of $\mb{H}$ and the uplink channel from user $k$, the columns of $\mb{A}_k \in \mathbb{C}^{N \times L}$ represent the array response for $L$ propagation paths, $\g_k \sim \mc{CN}(\mb{0},\mb{I}_L)$ represents the small-scale fading, and $\beta_k$ models the geometric attenuation and slow fading. We assume that $\mb{A}_k$ is a full rank matrix whose $\ell$-th column is the array steering vector corresponding to the angle of arrival (AoA) $\theta_{k\ell}$ of the $\ell$-th path, as given by
\begin{align} \label{eq:steering_vector}
    \mb{a} (\theta_{k\ell}) = \big[1,\e^{-j2\pi \frac{d}{\lambda} \sin\theta_{k\ell}},\ldots, \e^{-j(N-1)2\pi \frac{d}{\lambda} \sin\theta_{k\ell}}\big]^T,
\end{align}
where $d$ is the inter-antenna spacing of the ULA and $\lambda$ denotes the wavelength. For simplicity, we denote $\omega_{{k\ell}} = 2\pi \frac{d}{\lambda} \sin\theta_{k\ell}$ as the spatial frequency of the $\ell$-th path from user $k$. 

We assume that the AoAs for all users are situated within a specific angular sector $\mc{S}_{\theta_0} \triangleq [\theta_0-\Theta/2 ,\theta_0 + \Theta/2]$, where $\theta_0$ is the center angle of the sector and $\Theta$ is the azimuth angular spread. 
This constraint frequently arises in practice because cell sectoring limits service to users within a specific angular sector of the BS coverage. Additionally, the high frequency and directional nature of mmWave signals produce narrow beam patterns and substantial attenuation of distant reflections, making nearby reflections of users more dominant and resulting in narrower angular spreads for users.
 {Note that the channel model considered in \eqref{eq:h_k} is narrowband, meaning the signal bandwidth is smaller than the coherence bandwidth. This model is relevant to applications such as short-range indoor communications \cite{zhang2022mmwave} and low-mobility IoT networks \cite{hussain2022integrated}, where LOS paths or narrow angular sectors dominate over frequency selectivity. This work leverages variational Bayesian inference to develop novel detection algorithms tailored for such scenarios.}
\vspace{0em}
\subsubsection{Mutual Coupling}
To take mutual coupling between antennas into account, the channel vector from user $k$ to the BS can be modeled as
\begin{align}
    \bar{\mb{h}}_k = \sqrt{{\beta_k}/{L}} \mb{T}\mb{A}_k \mb{g}_k,
\end{align}
where the coupling matrix $\mb{T}\in \mathbb{C}^{N \times N}$ is given by
\begin{align}\label{eq:mc_mat}
	\mb{T} = \big(\mb{I} + R^{-1}\mb{Z} \big)^{-1}
\end{align}
and $R$ denotes the low-noise amplifier (LNA) input impedance. Assuming thin half-wavelength dipoles, the elements of $\mb{Z}$ are expressed as \cite{pirzadehmc,rao2021massive}
\begin{align}
\label{eq:mutual_coupling}
	&[\mb{Z}]_{ij} = 30\big(2\mathrm{Ci}(2\pi d_{ij}) - \mathrm{Ci}(\xi_{ij}+\pi) - \mathrm{Ci}(\xi_{ij} - \pi)\nonumber\\
	&\quad + j\left(-2\mathrm{Si}(2\pi d_{ij}) + \mathrm{Si}(\xi_{ij}+\pi) + \mathrm{Si}(\xi_{ij} - \pi)\right)
	\big),~i\neq j\nonumber\\
&[\mb{Z}]_{ii} = 30\bigl(\gamma +\mathrm{log}(2\pi)-\mathrm{Ci}(2\pi)+j\mathrm{Si}(2\pi)\bigr),	
\end{align}
where $d_{ij}$ denotes the distance between antennas $i$ and $j$ normalized by the wavelength, and $\xi_{ij}=\pi\sqrt{1+4d_{ij}^{2}}$.\\
\indent Mutual coupling between antenna elements influences not only the channel characteristics but also the statistical properties of the noise at the receiver. Unlike traditional quantization schemes, the $\sd$ architecture leverages spatial oversampling or users confined to narrow angular regions to shape the quantization noise away from desired signal directions. However, mutual coupling introduces colored noise, denoted as $\n_{\rm c} \sim \mc{CN}(\mb{0}, \mb{R})$, with covariance matrix $\mb{R}$. When considering $\sd$-based MIMO systems with less than half-wavelength-spaced antennas, it is important to consider the potential impact of mutual coupling. 
Based on the above electromagnetic coupling model, $\mb{R}$ can be expressed as \cite{pirzadehmc,rao2021massive}
\begin{align}\label{eq:R}
\mb{R} = \mb{T}\big(\sigma_i^2(\mb{Z}\mb{Z}^H + R_c^2\mb{I} - 2R_c\mathfrak{R}(\rho^*\mb{Z})) + 4\mathrm{k}TB\mathfrak{R}(\mb{Z})\big)\mb{T}^H,
\end{align}
where $\mathbb{E}\{\mb{i}_c\mb{i}_c^H\}=\sigma_i^2\mb{I}$, $\mathbb{E}\{\mb{u}_c\mb{u}_c^H\}=\sigma_u^2\mb{I}$, $R_c = \frac{\sigma_u}{\sigma_i}$, $\mathbb{E}\{\mb{u}_c\mb{i}_c^H\}=\frac{\rho}{\sigma_u\sigma_i}\mb{I}$, $\mb{i}_c$ and $\mb{u}_c$ represent the equivalent noise current and voltage of the LNA, and $\mathrm{k}$, $T$, and $B$ denote the Boltzmann constant, environment temperature, and bandwidth, respectively.  
The received analog signals are then quantized using few-bit ADCs. The mathematical models for both conventional and $\sd$ quantizers in these few-bit ADCs will be presented in the following subsection.
%
%

%
\subsection{MIMO Receiver with Few-Bit ADCs}
\subsubsection{Few-Bit ADCs with Conventional Quantizers}
The received analog signal at each BS antenna is sampled to produce a discrete-time signal, which is then quantized by a pair of $b$-bit ADCs to one of a set of discrete quantized levels. Thus, the quantized signal $\mb{y}= \mc{Q}_b(\mb{x})$ can be modeled as
\begin{align}
    \mathfrak{R}\{\mb{y}\} = \mc{Q}_b(\mathfrak{R}\{\mb{x}\}), \qquad  \mathfrak{I}\{\mb{y}\} = \mc{Q}_b(\mathfrak{I}\{\mb{x}\}),
\end{align}
where $\mc{Q}_b(.)$ denotes the $b$-bit ADC operation applied element-wise to its vector argument. The uniform quantizer $\mc{Q}_b(.)$ performs $b$-bit scalar quantization, which produces $2^b-1$ quantized output levels belonging to the set $\{d_1,\ldots,d_{2^b-1}\}$. Without loss of generality, we assume ${-\infty = d_0 <d_1\ldots <d_{2^b-1} <d_{2^b} = +\infty}$. The quantization thresholds are given by \cite{nguyen2023variational}
\begin{align}
    d_m \!=\! \big(-2^{b-1} + m\big)\Lambda, \quad \forall m \in \mathcal{M} = \{1,\ldots,2^{b}-1\},
\end{align}
where $\Lambda$ denotes the quantization step size of $\mc{Q}_b(.)$. The quantized output $q_i$ at the $i$-th antenna of the BS is defined as
\begin{align}
    q_i = \mc{Q}_b(x_i) = 
    \begin{cases}
        d_m - \frac{\Lambda}{2}, & {\rm if} \, x_i \in (d_{m-1},d_{m}), \\
		(2^b-1)\frac{\Lambda}{2}, & {\rm if}\, x_i \in (d_{2^b-1},d_{2^b}).
    \end{cases}
\end{align}
We also define by $q_i^\up$ and $q_i^\low$ the upper and lower thresholds of the quantization bin to which $q_i$ belongs.
\subsubsection{Few-Bit ADCs with Sigma-Delta Quantizers}
\begin{figure}[t]
	\centering
 \includegraphics[width=.7\linewidth]{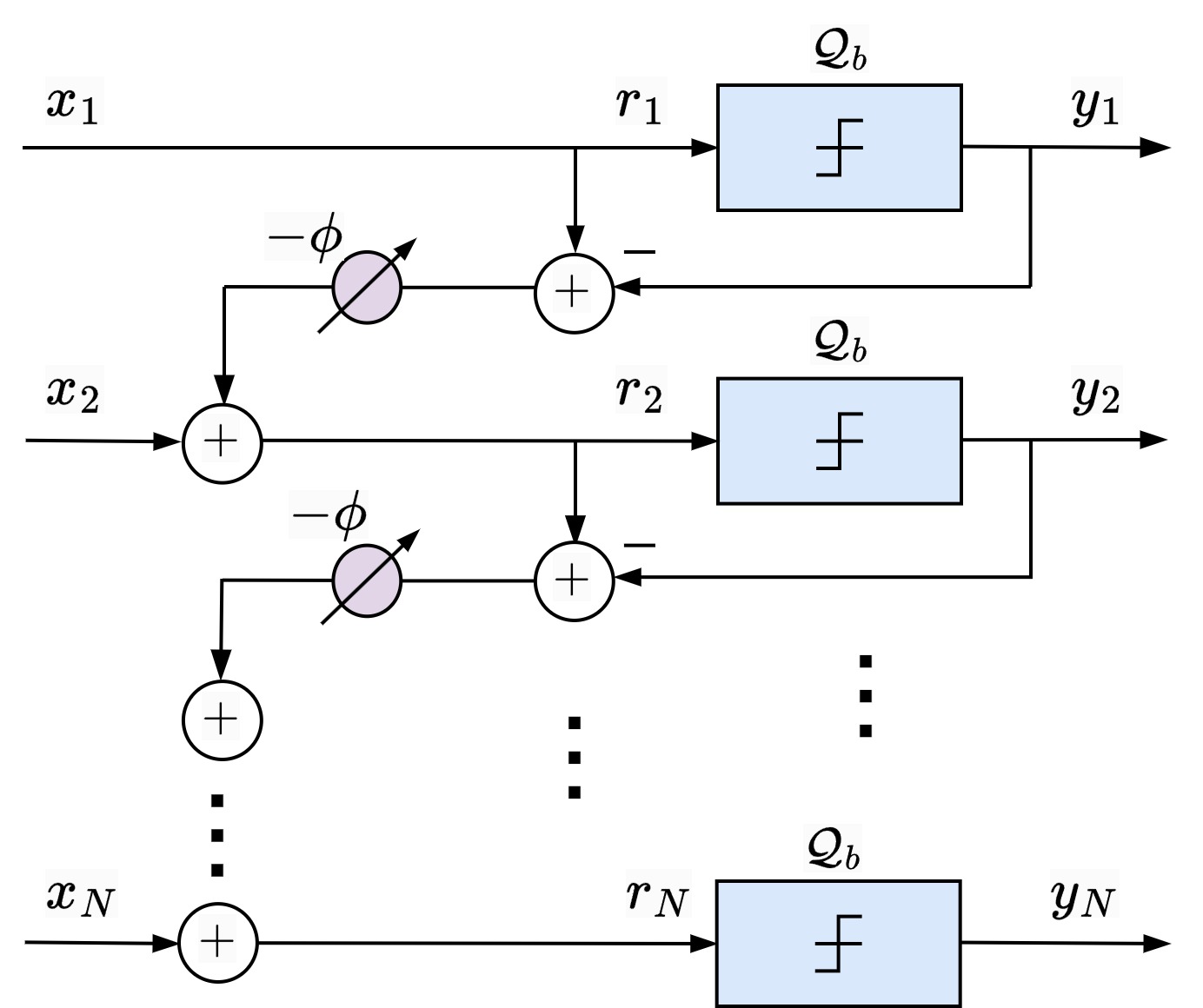}
	\caption{Illustration of the spatial $1^{\st}$-order $\sd$ architecture at an $N$-antenna receiver.}
	\label{1stOrderSigmaDelta}
\end{figure}

The $\Sigma\Delta$ quantization process is illustrated in Fig.~\ref{1stOrderSigmaDelta}, 
where a few-bit quantizer is used in each quantization block. The pre-quantized signal at stage $i$, denoted by $r_i$, which consists of the unquantized received signal $x_i$ and the difference between the input and output of the previous quantizer, can be represented as \cite{rao2021massive,pirzadeh2020spectral}
\begin{align} \label{eq:quan:signal}
r_i = x_i + \e^{-j\phi} (r_{i-1} - y_{i-1}), \forall i = 1,\ldots, N,
\end{align}
where $x_i \triangleq {\bf h}_i^T {\bf s} +n_i$ is the unquantized received signal at antenna $i$, $r_0$ and $y_0$ are set to $0$, and $\phi$ denotes a phase shift applied to the signal between two adjacent antennas.
After quantization, the observation at the $i$-th antenna element can be expressed as 
\begin{align} \label{eq:quan:obsevation}
    y_i = \mathcal{Q}_b(r_i).
\end{align} 


%

Using spatial oversampling with antenna elements spaced closer than one-half wavelength, and feedback of the quantization error between adjacent antennas, the $\sd$ architecture shapes the quantization noise to higher spatial frequencies, thereby significantly reducing it for signals arriving from angles closer to the broadside of the ULA (when $\phi=0$). However, there is a practical limit to using spatial oversampling due to the physical dimensions of the antennas and mutual coupling. For this reason, the $\sd$ architecture is best suited for scenarios where two conditions are met: the array incorporates spatial oversampling and the users of interest are confined within a specific angular sector \cite{rao2021massive}.

In the following, we present prior art on the linearized model for spatial $1^{\st}$-order $\sd$ ADCs and corresponding linear minimum mean-squared error (LMMSE) receiver to estimate $\mb{s}$. Subsequent developments will include the design of efficient VB-based MIMO detection algorithms for $1^{\st}$- and $2^{\nd}$-order $\sd$ ADCs.
\subsection{Linear Detection with Spatial $\Sigma\Delta$ Quantized Observations} \label{sec:LMMSE}
Using the Bussgang decomposition \cite{bussgang1952crosscorrelation}, the output of the $\sd$ array in \eqref{eq:quan:obsevation} in vector form is linearized as 
\begin{align}\label{eq:linear:y:1}
 {\mb{y} = \mb{x} + \mb{U}^{-1}\mb{q},}
\end{align}
where $\mb{q}= [q_1,\ldots,q_N]^T$ represents the effective quantization noise, {and $\mb{U}$ is given in \eqref{eq:matrix:U}. A detailed derivation of \eqref{eq:linear:y:1} is provided in Appendix \ref{appendix-0}. In \eqref{eq:linear:y:1}, the unquantized signal $\mb{x}$ is separated from the shaped quantization noise $\mb{U}^{-1}\mb{q}$, which enables the use of linear detection methods for the $\sd$ array. 
}
For the special case of one-bit $\sd$ quantization, the variance of $q_i$ is given by \cite[Eq. (33)]{pirzadeh2020spectral}
\begin{align} \label{eq:var:qi}
    \tau_{q_i} = \Big(\frac{\pi}{2} -1\Big)\frac{1-(\pi/2-1)^i}{2-\pi/2}p_s,
\end{align}
where $p_s$ is the power of the received signal at the $i$-th antenna.

Based on the linearized model in \eqref{eq:linear:y:1}, an LMMSE receiver $\mb{W}$ that minimizes the mean squared error ${\mathbb{E}\big[\|\mb{s} - \mb{Wy} \|^2 \big]}$ in estimating $\mb{s}$ can be expressed as
\begin{align} \label{LMMSE:detector}
    \mb{W} = \bs{\Sigma}_\mb{s}\mb{H}^H(\mb{H}\bs{\Sigma}_\mb{s}\mb{H}^H + N_0 \mb{I}_N + \mb{U}^{-1}\bs{\Sigma}_{\mb{q}} \mb{U}^{-H})^{-1},
\end{align}
where $\bs{\Sigma}_\mb{s}$ 
and $\bs{\Sigma}_{\mb{q}} = \textrm{diag}(\tau_{q_1},\ldots,\tau_{q_N})$ are the covariance matrices of $\mb{s}$ and $\mb{q}$, respectively. 

 {
If we take the mutual coupling effect into account, the LMMSE solution for estimating $\mb{s}$ can be obtained as
\begin{align} \label{LMMSE:detector:MC}
    \mb{W}^{\rm MC} = \bs{\Sigma}_\mb{s}\mb{H}^H(\mb{H}\bs{\Sigma}_\mb{s}\mb{H}^H + \mb{R} + \mb{U}^{-1}\bs{\Sigma}^{\rm MC}_{\mb{q}} \mb{U}^{-H})^{-1} \mb{y} ,
\end{align}
where the covariance matrix of the quantization noise $\bs{\Sigma}^{\rm MC}_{\mb{q}}$ can be approximated as \cite{pirzadeh2020effect}
\begin{align} 
\bs{\Sigma}^{\rm MC}_{\mb{q}} \simeq \mathrm{diag}(\mb{p}_{\mb{q}}),
\end{align}
and 
\begin{eqnarray}
    \mb{p}_{\mb{q}} & = & \left(\frac{\pi}{2}\zeta-1\right)\bs{\Pi}\mb{p}_{\mb{x}}, \\
\mb{p}_{\mb{x}} & = & 
\begin{bmatrix}
        \mathbb{E}\big[|x_1|^2\big], \mathbb{E}\big[|x_2|^2\big],\cdots,\mathbb{E}\big[|x_N|^2\big]
\end{bmatrix}^T,
\end{eqnarray}
\begin{align}
&\bf{\Pi}=\nonumber\\
 &\begin{bmatrix}
    1 & 0 & \ldots & & 0 & 0 \cr
    \left(\frac{\pi}{2}\zeta-1\right) & 1 & 0& \ldots& & 0 \cr
    \vdots & \ddots & 1 & & & \vdots \cr \left(\frac{\pi}{2}\zeta-1\right)^{m} & \ddots& \ddots & \ddots & & \cr \vdots& \ddots & \ddots& \ddots & \ddots & \cr \left(\frac{\pi}{2}\zeta-1\right)^{M-1} &\cdots & \left(\frac{\pi}{2}\zeta-1\right)^{m}& \cdots&\left(\frac{\pi}{2}\zeta-1\right) & 1
    \end{bmatrix},
\end{align}
where $\zeta=1.13$ is a correction factor.
}

As mentioned earlier, the accuracy of the Bussgang decomposition depends on the Gaussianity of the input to the quantization non-linearity, which does not always hold, particularly when the input symbol is discrete. To overcome this challenge, in the next section, we introduce an alternative MIMO detection approach based on VB inference to handle the nonlinear $\sd$ receiver architecture. 

\section{Variational Bayes for Few-Bit $1^{\st}$-Order $\Sigma\Delta$ MIMO Detection} \label{sec:1stSD}
In this section, we first provide background on VB inference and then develop a new data detection method based on the VB framework in few-bit $\Sigma\Delta$ MIMO systems with known channel $\mb{H}$.
\vspace{-0.2cm}
\subsection{Background on VB Inference} \label{subsec:VB}
We denote the set of all observed variables as $\bar{\mb{y}}$ and the set of $V$ latent variables as $\bar{\mb{x}}$. To find a Bayes estimate of $\bar{\mb{x}}$, one would need to determine the posterior distribution $p(\bar{\mb{x}}|\bar{\mb{y}})$, which is often computationally intractable. To overcome this challenge, the VB method aims to find a distribution $q(\bar{\mb{x}})$ characterized by variational parameters within a predefined family $\mc{Q}$ of densities, such that $q(\bar{\mb{x}})$ closely approximates $p(\bar{\mb{x}}|\bar{\mb{y}})$. To this end, VB defines an optimization problem leveraging the Kullback-Leibler (KL) divergence from $q(\bar{\mb{x}})$ to $p(\bar{\mb{x}}|\bar{\mb{y}})$:
\begin{align}
\label{eq:opt_KL}
q^{\star}(\bar{\mb{x}}) =& \arg\min_{q(\bar{\mb{x}}) \in \mc{Q}} \text{KL}\big(q(\bar{\mb{x}}) \| p(\bar{\mb{x}}|\bar{\mb{y}}) \big),\\
=& \arg\min_{q(\bar{\mb{x}}) \in \mc{Q}} \big\{\mathbb{E}_{q(\bar{\mb{x}})} \big[\ln q(\bar{\mb{x}}) \big] - \mathbb{E}_{q(\bar{\mb{x}})} \big[\ln p(\bar{\mb{x}}|\bar{\mb{y}})\big] \big\}, \nonumber
\end{align}
where $q^{\star}(\bar{\mb{x}})$ denotes the optimal variational distribution. 
 {To approximate the posterior, we adopt the mean-field VB inference method, for which the variational distribution $q(\bar{\mb{x}})$ is drawn from a mean-field variational family \cite{bishop2006pattern}, such that}
\begin{equation}
q(\bar{\mb{x}}) = \prod_{i=1}^V q_i(x_i),
\end{equation}
where the latent variables are mutually independent, with each being governed by a distinct factor in the variational density. The optimal value of $q_i(x_i)$ is obtained as \cite[Chapter 10]{bishop2006pattern}
\begin{align}
 \label{eq:q_start_prop}
		q_i^{\star}(x_i) &\propto \mr{exp}\left\{\big\langle{\ln p (\bar{\mb{y}},\bar{\mb{x}})\big\rangle}\right\},
\end{align}
 {where $\lr{\cdot}$ is computed utilizing the currently fixed variational density $q_{-i}(\bar{\mb{x}}_{-i}) = \prod_{j=1, j\neq i}^{V} q_{j}(x_{j})$}. To optimize~\eqref{eq:opt_KL}, we use the Coordinate Ascent Variational Inference algorithm, which is an iterative method that ensures convergence to at least a locally optimal solution \cite{wainwright2008graphical}. 
\subsection{VB Inference for Few-bit $\sd$ MIMO Detection}
Considering the input-output relationship of the $1^{\st}$-order $\Sigma\Delta$ quantizer in \eqref{eq:quan:signal}, the conditional distributions between the observed quantized signal $y_i$ and latent variables, i.e., $r_i$, $x_i$, $\bf s$, are given by $p(r_i|r_{i-1},y_{i-1}, \mb{s};\mb{H},N_0) = \mathcal{CN}(r_i;\mb{h}_i^T\mb{s} + \e^{-j\phi}(r_{i-1} - y_{i-1}),N_0)$ and $p(y_i|r_i) = \mathbbm{1}(r_i \in [y_i^\low,y_i^\up])$, where $\mathbbm{1}$ is the indicator function which equals one if the argument holds, or zero otherwise.
The optimal MAP detector $\hat{\bf s}_{\tt MAP} = \arg \max_{{\bf s} \in \mathcal{S}^K} p(\mb{y},\mb{s};\mb{H},N_0)$ with spatial $\Sigma\Delta$ ADCs can be written as
\begin{align} \label{MAP:es}
\hat{\bf s}_{\tt MAP} &= \arg \max_{{\bf s} \in \mathcal{S}^K} \int p(\mb{y}, \mb{r}, \mb{s};\mb{H},N_0) \mr{d} \mb{r}  \\
&= \!\arg \!\max_{{\bf s} \in \mathcal{S}^K} \,\!p(\mb{s})\!\int \! \prod_{i=1}^N\! p(y_i|r_i) p(r_i|r_{i-1},y_{i-1}, \mb{s};\mb{H},N_0) \mr{d} \mb{r}. \nonumber
\end{align}
This integral is intractable to evaluate in closed-form due to the high-dimensional integration and the intricate correlation between $\mb{r}$ and $\mb{s}$. This challenge motivates us to explore a novel VB inference method to solve the data detection problem with $\Sigma\Delta$ arrays. 
\begin{figure}[!t]
    \centering
    \includegraphics[width=0.45\linewidth]{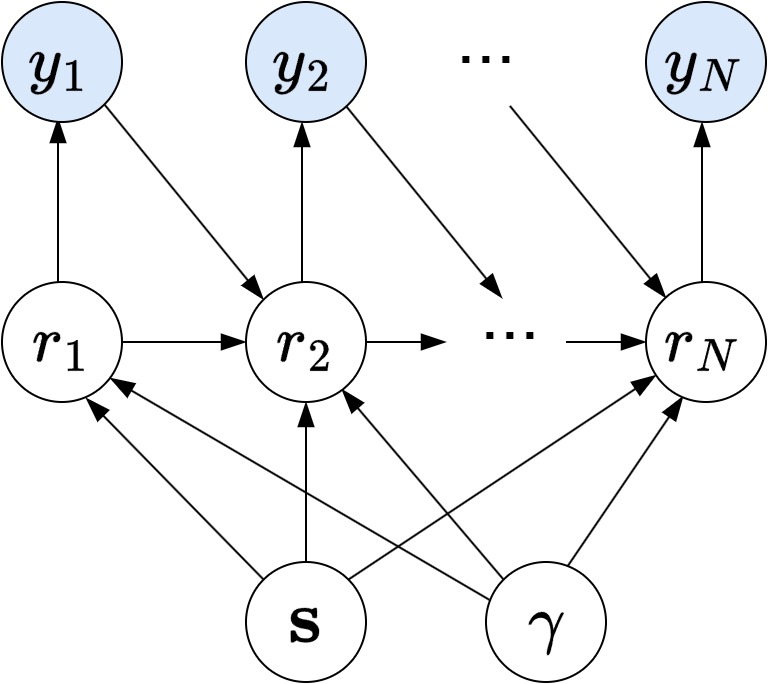}
    \caption{The Bayesian network for the $1^{\st}$-order $\sd$ receiver.}
    \label{fig:Bayesian-Sigma-Delta-1st}
\end{figure}

We consider the residual interference-plus-noise as an unknown parameter $N_0^{\rm post}$, which must be estimated using the VB method. The dependency between random variables under spatial $\Sigma\Delta$ processing can be graphically modeled through the Bayesian network in Fig.~\ref{fig:Bayesian-Sigma-Delta-1st}, where $\gamma \triangleq 1/N_0^{\rm post}$ denotes the precision that is floated as an unknown random variable and the arrows indicate the conditional probability between variables.
The main goal is to infer the distribution of data $\mb{s}$ given the observation $\mb{y}$. 
To accomplish this, we employ the mean-field VB inference method, which derives the mean-field variational distributions $q(r_i)$, $q(s_k)$, and $q(\gamma)$, from the factorized family to approximate the posterior for individual variables, such that
\begin{align} \label{eq:mean-field}
p(\mb{s},\mb{r},\gamma|\mb{y};\mb{H}) \approx q(\mb{s},\mb{r},\gamma) = \prod_{k=1}^K q(s_k) \prod_{i=1}^N q(r_i)q(\gamma).
\end{align}
According to \eqref{eq:q_start_prop}, to obtain the optimal solution of the variational densities in \eqref{eq:mean-field}, we need the joint distribution $p(\mb{y},\mb{r},\mb{s},\gamma;\mb{H})$ which can be factorized as
\begin{align} \label{joint:dist}
  p(\mb{y},\mb{r},\mb{s},\gamma;\mb{H}) &= p(\mb{s}) p(\gamma) \nonumber \\
  &\;\;\times \prod_{i=1}^N p(y_i|r_i) p(r_i|r_{i-1},y_{i-1},\mb{s},\gamma;\mb{H}),
\end{align}
where 
\begin{align}
    &p(r_i|r_{i-1},y_{i-1}, \mb{s},\gamma;\mb{H}) \nonumber \\
    &= \mathcal{CN}\big(r_i;\mb{h}_i^T\mb{s} + \e^{-j\phi}(r_{i-1} - y_{i-1}),\gamma^{-1}\big).
\end{align}
To this end, we present the update process for each latent variable through the derived closed-form expression of each variational density.

\textit{1) Updating $r_i$:} For $i=1,\ldots,N-1$, the variational distribution $q(r_i)$ can be obtained by taking the expectation of the conditional in \eqref{joint:dist} w.r.t. $q(\mb{s},\gamma)$ as
\begin{align} \label{q:r_i:1st}
  q(r_i) &\propto \exp\Big\{\blr{\ln p(y_i|r_i) + \ln p(r_{i+1}|r_i,y_i,\mb{s},\gamma;\mb{H}) \nonumber\\
  &\qquad +\ln p(r_i|r_{i-1},y_{i-1},\mb{s},\gamma;\mb{H})}\Big\}\nonumber\\
  &\propto p(y_i|r_i) \exp\!\Big\{\! -\! \blr{\gamma |r_{i} - \mb{h}_{i}^T\mb{s} - \e^{-j\phi}(r_{i-1} - y_{i-1})|^2}  \nonumber\\
  &\hskip7em\!\! -\! \blr{ \gamma |r_{i+1} - \mb{h}_{i+1}^T\mb{s} - \e^{-j\phi}(r_{i}  - y_{i})|^2} \Big\}\nonumber\\
&\propto p(y_i|r_i) \exp\big\{ - \langle \gamma \rangle \big( |r_i  - a_i|^2 + |r_{i} - b_i |^2\big)\big\} \nonumber\\
&\propto \mathbbm{1}(r_i \in [y_i^\low,y_i^\up])\, \mathcal{CN}\big(r_i;(a_i+b_i)/2,1/(2\langle \gamma \rangle) \big),
\end{align} 
where $a_i \triangleq \mb{h}_{i}^T \langle \mb{s}\rangle  + \e^{-j\phi}(\langle r_{i-1} \rangle - y_{i-1})$ and $b_i \triangleq y_i + \e^{j\phi}\langle r_{i+1} \rangle  - \e^{j\phi}\mb{h}_{i+1}^T \langle \mb{s}\rangle$.

Defining $v_i \triangleq (a_i+b_i)/2$, we see that the variational distribution $q(r_i)$ in \eqref{q:r_i:1st} 
is a truncated complex Gaussian distribution obtained by pruning the distribution $\mathcal{CN}\big(v_i,1/(2\lr{\gamma})\big)$ onto the interval $[y_i^\low,y_i^\up]$.
In Appendix \ref{appendix-1} we show that the mean and variance of the resulting distribution are given by
${\langle r_i \rangle = \ms{F}_r (v_i,2\langle \gamma \rangle,y_i^\low,y_i^\up)}$ and ${\tau_{r_i} = {\sf G}_r (v_i,2\langle \gamma \rangle,y_i^\low,y_i^\up)}$, respectively.

For $i=N$, while the factor $p(r_{N+1}|r_N,y_N,\mb{s},\gamma;\mb{H})$ does not exist, the variational distribution $q(r_N)$ can be computed similarly to \eqref{q:r_i:1st}
and is given by 
\begin{align} \label{q:r_N:1st}
  q(r_N) &\propto \mathbbm{1}(r_N \in [y_N^\low,y_N^\up])\, \mathcal{CN}\big(r_N;a_N,1/\langle \gamma \rangle \big).
\end{align}

\textit{2) Updating $s_k$:} The variational distribution $q(s_k)$ is obtained by taking the expectation of the conditional in \eqref{joint:dist} w.r.t. $q(\mb{r},\gamma)$:
\begin{align}
  q(s_k) &\propto \exp\bigg\{\!\bblr{\ln p(s_k) + \sum_{i=1}^{N}\ln p(r_{i}|r_{i-1},y_{i-1},\mb{s},\gamma;\mb{H})}\! \bigg\}\nonumber\\
&\propto p(s_k) \exp\bigg\{\!-\lr{\gamma}\sum_{i=1}^{N} \blr{|{r_i} - \mb{h}_{i}^T\mb{s} \nonumber\\
&\hskip9em - \e^{-j\phi}({ r_{i-1} }  - y_{i-1})|^2} \bigg\}.\!
\end{align}
Expanding $\mb{h}_{i}^T  \mb{s} = {h}_{i,k}s_k + \sum_{j\neq k}^K {h}_{i,j}  s_j $ leads to
\vspace{-0.15cm}
\begin{align} \label{q:s_k}
  q(s_k) 
  &\propto p(s_k) \exp\bigg\{ -\lr{ \gamma } \sum_{i=1}^{N} \bigg|\lr{ r_{i}} - \sum_{j\neq k}^K {h}_{i,j}  \lr{ s_j}\nonumber\\
  &\hskip7em- \e^{-j\phi}(\lr{r_{i-1}}  - y_{i-1}) - h_{i,k}s_k \bigg|^2 \bigg\}\nonumber\\
  &\propto p(s_k) \exp\bigg\{ -\lr{\gamma } \sum_{i=1}^{N} |z_{i,k} - {h}_{i,k}s_k|^2 \bigg\}\nonumber\\
& \propto p(s_k) \prod_{i=1}^N \mathcal{CN} \big(z_{i,k};{h}_{i,k}s_k, \lr{ \gamma } ^{-1} \big),
\end{align}
where
\begin{align}\label{eq:z_ik}
    z_{i,k} &= \!\lr{ r_{i}} - \e^{-j\phi}(\lr{r_{i-1}}  - y_{i-1}) - \sum_{j\neq i}^K {h}_{i,j}  \lr{ s_j } \nonumber\\
    &=\! \lr{ r_{i}} - \e^{-j\phi}(\lr{ r_{i-1}}  - y_{i-1}) - \mb{h}_{i}^T \lr{ \mb{s} } + h_{i,k} \lr{ s_k },
\end{align}
using the current estimate $\lr{s_k}$ for all $k=1,\ldots,K$. 
Interestingly, the variational distribution $q(s_k)$ is equivalent to the posterior distribution $p(s_k|\mb{z}_k,\lr{\gamma};\mb{h}_k)$ of $s_k$ in a single-input multiple-output (SIMO) system:
{
\begin{align} \label{eq:zk:SIMO}
  \mb{z}_k = \mb{h}_k s_k  + \Tilde{\mb{n}}_k, 
\end{align}
where $\Tilde{\mb{n}}_k \sim \mathcal{CN}(\mb{0}, \lr{\gamma} ^{-1} \mb{I}_N)$} and $\mb{z}_k = [z_{1,k},\ldots,z_{N,k}]^T$. The variational mean and variance of $s_k$ can be expressed as $\lr{ s_k }  = {\sf F}_s(\mb{z}_k,\mb{h}_k,\lr{\gamma})$ and $\tau_{s_k} = {\sf G}_s(\mb{z}_k,\mb{h}_k,\lr{\gamma})$, respectively, as shown in Appendix \ref{appendix-2}.

\textit{3) Updating $\gamma$:} 
 {We assume a Gamma distribution $p(\gamma) = {\rm Gamma}(\alpha,\beta)$, as a conjugate prior for the precision parameter $\gamma$, where $\alpha$ and $\beta$ are the shape and rate parameters, respectively. This enables the derivation of the variational distribution $q(\gamma)$ in closed form. The choice of a Gamma distribution for $\gamma$ is motivated by its role as a conjugate prior for the precision of a Gaussian likelihood in the noise model \eqref{eq:x}, and this conjugate prior is proportional to the product of a power of $\gamma$ and the exponential of a linear function of $\gamma$ \cite[Chapter 2, Section 2.3.6]{bishop2006pattern}.
Taking the expectation of the conditional distribution in \eqref{joint:dist} w.r.t. $q(\mb{s}, \mb{r})$, $q(\gamma)$ can be obtained as}
\begin{align}
q(\gamma) &\propto \exp\bigg\{\bblr{\sum_{i=1}^N\ln p(r_i|r_{i-1},y_{i-1},\mb{s},\gamma;\mb{H}) + \ln p(\gamma)}\bigg\}\nonumber\\
&\propto \exp \bigg\{ -  \gamma \sum_{i=1}^N \blr{| r_{i}  - \mb{h}_{i}^T  \mb{s} - \e^{-j\phi}( r_{i-1} - y_{i-1})|^2} \nonumber\\
&\hskip3em + N \ln \gamma + (\alpha - 1)\ln \gamma -\beta \gamma \bigg\}\nonumber\\
&\propto \exp \bigg\{-  \gamma  \sum_{i=1}^N \big( \blr{|r_i - \mb{h}_i^T\mb{s}|^2} + \blr{|r_{i-1} - y_{i-1}|^2} \nonumber\\
&\hskip4em- 2\,\Re\big\{\blr{(r_i - \mb{h}_i^T \mb{s})^*(r_{i-1} - y_{i-1})\e^{-j\phi}} \big\}\big) \nonumber\\
&\hskip4em+ (N + \alpha - 1)\ln \gamma -\beta \gamma \bigg\}.
 \end{align}
Using the expansion
 $\blr{|r_i - \mb{h}_i^T\mb{s}|^2}  = |\langle r_i \rangle - \mb{h}_i^T \langle\mb{s}\rangle |^2 + \tau_{r_i} + \mb{h}_i^T \bs{\Sigma}_{\mb{s}} \mb{h}_i$, we arrive at

\begingroup
\allowdisplaybreaks
 \begin{align}
q(\gamma) &\propto \exp \bigg\{\!\!-\gamma  \sum_{i=1}^N \!\Big[ |\langle r_i \rangle - \mb{h}_i^T \langle\mb{s}\rangle -  ( \langle r_{i-1} \rangle - y_{i-1})\e^{-j\phi} |^2 \nonumber\\
&\quad+ \tau_{r_i}\! + \mb{h}_i^H \bs{\Sigma}_{\mb{s}} \mb{h}_i + \tau_{r_{i-1}} \Big]\! + (N + \alpha - 1)\ln \gamma -\beta \gamma \bigg\}\nonumber\\
&\propto \exp \bigg\{-  \gamma  \sum_{i=1}^N \Big[ |u_i|^2 + \tau_{r_i} + \mb{h}_i^H \bs{\Sigma}_{\mb{s}} \mb{h}_i + \tau_{r_{i-1}} \Big] \nonumber\\
&\quad+ (N + \alpha - 1)\ln \gamma -\beta \gamma \bigg\} \nonumber\\
&\propto \exp \Big\{-  \gamma \Big[ \beta + \|\mb{u}\|^2 + 2 \tr\{\bs{\Sigma}_\mb{r}\} - \tau_{r_N}   \nonumber\\
&\quad + \tr \big\{ \mb{H}\bs{\Sigma}_{\mb{s}}\mb{H}^H\big\} \Big] + (N + \alpha - 1)\ln \gamma \Big\},
\end{align}
\endgroup
where $u_{i} \triangleq \lr{r_{i}} - \mb{h}_{i}^T \lr{\mb{s}}  - \e^{-j\phi}(\lr{ r_{i-1}} - y_{i-1})$, $\mb{u} = [u_1,\ldots,u_N]^T$ and $\bs{\Sigma}_\mb{r} = {\rm diag}(\tau_{r_1},\ldots,\tau_{r_N}) $ is the covariance matrix of $\mb{r}$. 
Here, $u_i$ denotes the residual term at receive antenna $i$, which would equals the noise term $n_i$ if $r_i$, $r_{i-1}$, and $\mb{s}$ were perfectly known.

 {Given that we have assigned a conjugate Gamma prior $p(\gamma) = {\rm Gamma}(\alpha,\beta)$, the resulting variational distribution follows a Gamma distribution. Specifically, after computing the expectation and simplifying the exponent, $q(\gamma)$ is a Gamma distribution with mean}
\begin{align}
    \langle \gamma \rangle = \frac{N+ \alpha}{\beta + \|\mb{u}\|^2 + 2 \tr\{\bs{\Sigma}_\mb{r}\} - \tau_{r_N}  + \tr \big\{ \mb{H}\bs{\Sigma}_{\mb{s}}\mb{H}^H\big\}},
\end{align}
 {where $N+ \alpha$ reflects the updated shape parameter, and the denominator incorporates the residual noise $\|\mb{u}\|^2$, the trace of the variance of $\mb{r}$, and the channel-data covariance term, consistent with the noise model and dependencies in the $1^{\st}$-order $\sd$ architecture.}

 {
By iteratively optimizing $q(r_i)$, $q(s_k)$, and $q(\gamma)$ through the updates of ${\lr{x_i}}$, ${\lr{s_k}}$, and $\lr{\gamma}$, we derive the CAVI algorithm, as presented in Algorithm~\ref{algo-1}, for the $1^{\st}$-order $\sd$ quantization. We designate this algorithm as the $1^{\st}$-order SD-VB algorithm,}
where the parameter $\epsilon$ is the numerator of $\hat{\gamma}$ in the variational distributions of \eqref{q:r_i:1st} and \eqref{q:r_N:1st} for updating $r_i$ and $r_N$, respectively.

\emph{Remark 1:}  {In Algorithm~\ref{algo-1}, we denote the estimates of the variational means $\lr{r_i}, \lr{s_k}, \lr{\gamma}$ at iteration $t$ as $\hat{r}_i^t, \hat{s}_k^t, \hat{\gamma}^t$, respectively,} and each iteration includes a round of updating the estimates of $\mb{r}, \mb{s}$, and $\gamma$.  {These estimates, computed via closed-form updates, represent the mean values of $q(r_i)$, $q(s_k)$, and $q(\gamma)$, driving the algorithm toward convergence.} Recalling $u_i = \lr{r_i} - \e^{-j\phi}(\lr{r_{i-1}} - y_{i-1}) - \mb{h}_i^T\lr{\mb{s}}$, $a_i = \mb{h}_{i}^T \lr{\mb{s}}  + \e^{-j\phi}(\lr{r_{i-1}} - y_{i-1})$ and $b_i = y_i + \e^{j\phi}\lr{r_{i+1}}  - \e^{j\phi}\mb{h}_{i+1}^T \lr{ \mb{s}}$, $a_i$ and $b_i$ can be rewritten as $a_i = \hat{r}_i^t - u_i$ and  $b_i = \hat{r}_i^t + \e^{j\phi}u_{i+1}$ using the current estimates $\big\{\hat{r}_i^t\big\}$ and $\big\{\hat{\mb{s}}^t\big\}$,  {where $\{\hat{r}_i^t\}$ denotes the set of estimated pre-quantized signals $\hat{r}_i^t$, across all antennas ($i=1,...,N$) at iteration $t$.} Thus, we have
$$v_i^t = \left\{\begin{array}{ll}
      \frac{a_i + b_i}{2} = \hat{r}_i^t - \frac{u_i - \e^{j\phi}u_{i+1}}{2}, &\textrm{ for } i=1,\ldots,N-1, \\
      a_N = \hat{r}_N^t - u_N, &\textrm{ for } i= N,
\end{array}\right.
$$
in step~7 of Algorithm~\ref{algo-1}. 

\emph{Remark 2:} The residual noise term $\mb{u}$, initialized as $\mb{u}= \hat{\mb{r}}^1 - \mb{H} \hat{\mb{s}}^1$, is introduced in lines $14$, $15$, and $21$ of Algorithm~\ref{algo-1} to reduce the computational complexity. Due to the sequential nature of VB, $v_i^t$ and $\mb{z}_k^t$ are calculated using the updated values of $\hat{\mb{r}}$ and $\hat{\mb{s}}$. Instead of computing $u_{i}= \lr{r_{i}} - \mb{h}_{i}^T \lr{\mb{s}}  - \e^{-j\phi}(\lr{ r_{i-1}} - y_{i-1})$ for all elements of $\mb{u}$, incurring a complexity of $\mathcal{O}(NK)$, the current value of the residual noise term $\mb{u}$ is used. The update of $\mb{u}$ reflects any update for the estimates of $\hat{r}_i$ or $\hat{s}_k$, and only incurs a complexity of $\mathcal{O}(N)$.

\begin{algorithm}[t]
\small
	\caption{\normalsize -- VB Algorithm for MIMO Detection with $1^{\st}$-Order $\Sigma\Delta$ Quantization}
	\begin{algorithmic}[1] 
		\STATE \textbf{Input:}
		$\mb{y}, \mb{H}$
        \STATE \textbf{Output:}	$\hat{\mb{s}}$
		\STATE Initialize $\hat{r}_i^1 = y_i^1$, ${\tau}_{r_i}^1=0$, $\forall i$, 
  $\hat{s}_k^1=0$, ${\tau}_{s_k}^1= {\rm Var}_{p(s_k)}[s_k]$, $\forall k$, $\mb{u}= \hat{\mb{r}}^1 - \mb{H} \hat{\mb{s}}^1$.
        \FOR{$t=1,2,\ldots$}
        \STATE $\hat{\gamma}^t \leftarrow (N+\alpha)/\big(\beta + \|\mb{u}\|^2 + 2 \tr\{\bs{\Sigma}_\mb{r}\} - \tau_{r_N}^t +\tr \{\mb{H}\bs{\Sigma}_{\mb{s}}\mb{H}^H\}\big)$
         \FOR{$i=1,\ldots,N$}
         \IF{$i=N$}
         \STATE $v_i^t \leftarrow \hat{r}_N^t - u_N$, $\epsilon = 1$
         \ELSE
        \STATE $v_i^t \leftarrow \hat{r}_i^t - (u_i - \e^{j\phi}u_{i+1})/2$, $\epsilon = 2$
        \ENDIF
        \STATE $\hat{r}_i^{t+1} \leftarrow \ms{F}_r(v_i^t,\epsilon\hat{\gamma}^t,y_i^{\low},y_i^{\up})$
        \STATE ${\tau}_{r_i}^{t+1} \leftarrow \ms{G}_r(v_i^t,\epsilon\hat{\gamma}^t,y_i^{\low},y_i^{\up})$
        \STATE $u_i \leftarrow u_i - \hat{r}_i^t + \hat{r}_i^{t+1}$
        \STATE $u_{i+1} \leftarrow u_{i+1} + \e^{-j\phi} (\hat{r}_i^t - \hat{r}_i^{t+1})$ only for $i<N$
        \ENDFOR
        \FOR{$k=1,\ldots,K$}
        \STATE $\mb{z}_k^t \leftarrow  \mb{h}_k\hat{s}_k^t+ \mb{u}$
        \STATE $s_k^{t+1} \leftarrow \ms{F}_s(\mb{z}^t_k,\mb{h}_k,\hat{\gamma}^t)$
        \STATE $\tau_{s_k}^{t+1} \leftarrow \ms{G}_{s}(\mb{z}^t_k,\mb{h}_k,\hat{\gamma}^t)$
        \STATE $\mb{u} \leftarrow \mb{u} + \mb{h}_k(\hat{s}_k^t - \hat{s}_k^{t+1})$
        \ENDFOR
        \ENDFOR
        \STATE $\forall k: \hat{s}_k \leftarrow \arg \max_{a\in \mathcal{S}}p_a \mathcal{CN}\big(\mb{z}_{k}^t;\mb{h}_k a,(1/\hat{\gamma}^t)\mb{I}_M\big)$
	\end{algorithmic} 
    \label{algo-1}
\end{algorithm}

\vspace{-0.15cm}
\section{Extension to MIMO Detection with $2^{\nd}$-Order $\Sigma\Delta$ Quantization} \label{sec:2ndSD}
\begin{figure}[t]
	\centering
 \includegraphics[width=.7\linewidth]{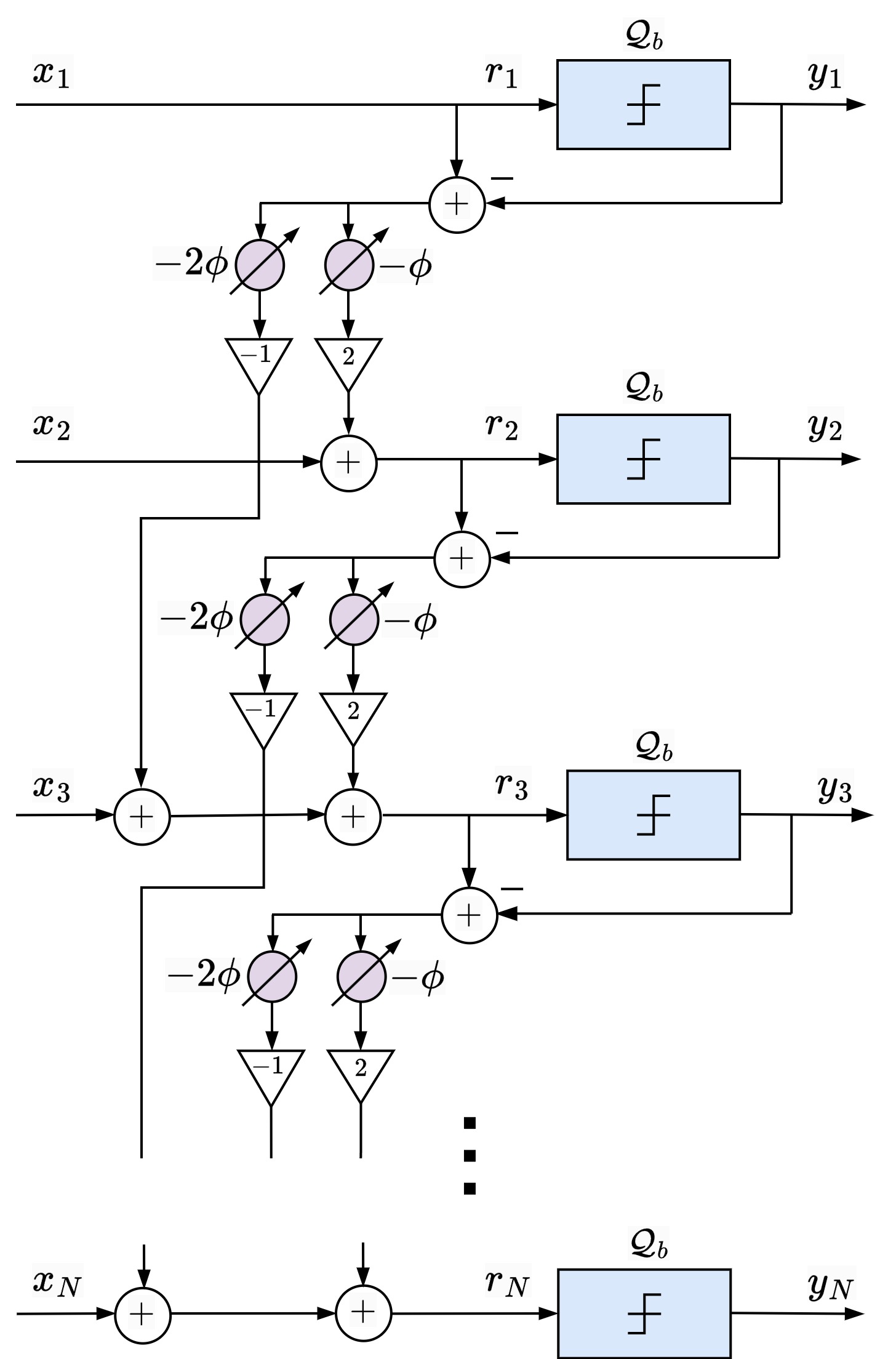}
	\caption{Illustration of the $2^{\nd}$-order spatial $\sd$ architecture for a multi-antenna receiver.}
	\label{2ndOrderSigmaDelta}
\end{figure}

The $2^{\nd}$-order $\sd$ architecture provides more aggressive noise shaping, further reducing the quantization noise in the desired angular sector compared to the $1^{\st}$-order case \cite{aziz1996overview}. 
In this section, we develop a VB approach for MIMO detection with $2^{\nd}$-order $\sd$ quantization based on the approach of the previous section. 
The $2^{\nd}$-order spatial $\sd$ architecture is illustrated in Fig.~\ref{2ndOrderSigmaDelta}. The phase difference applied between antennas $i$ and $(i+1)$ is $\phi$, while that between antenna $i$ and $(i+2)$ is $2\phi$.
The input-output relationship for the $2^{\nd}$-order $\sd$ approach is given by
\begin{align} \label{eq:inout}
    r_i &= x_i + 2\e^{-j\phi}(r_{i-1} - y_{i-1}) - \e^{-j2\phi}(r_{i-2} - y_{i-2}), \\
    y_i &= \mc{Q}_b(r_i).
\end{align} 

%
\vspace{-0.2cm}
\subsection{Proposed SD-VB Approach for Data Detection}
\begin{figure}[!t]
	\centering
 \includegraphics[width=.75\linewidth]{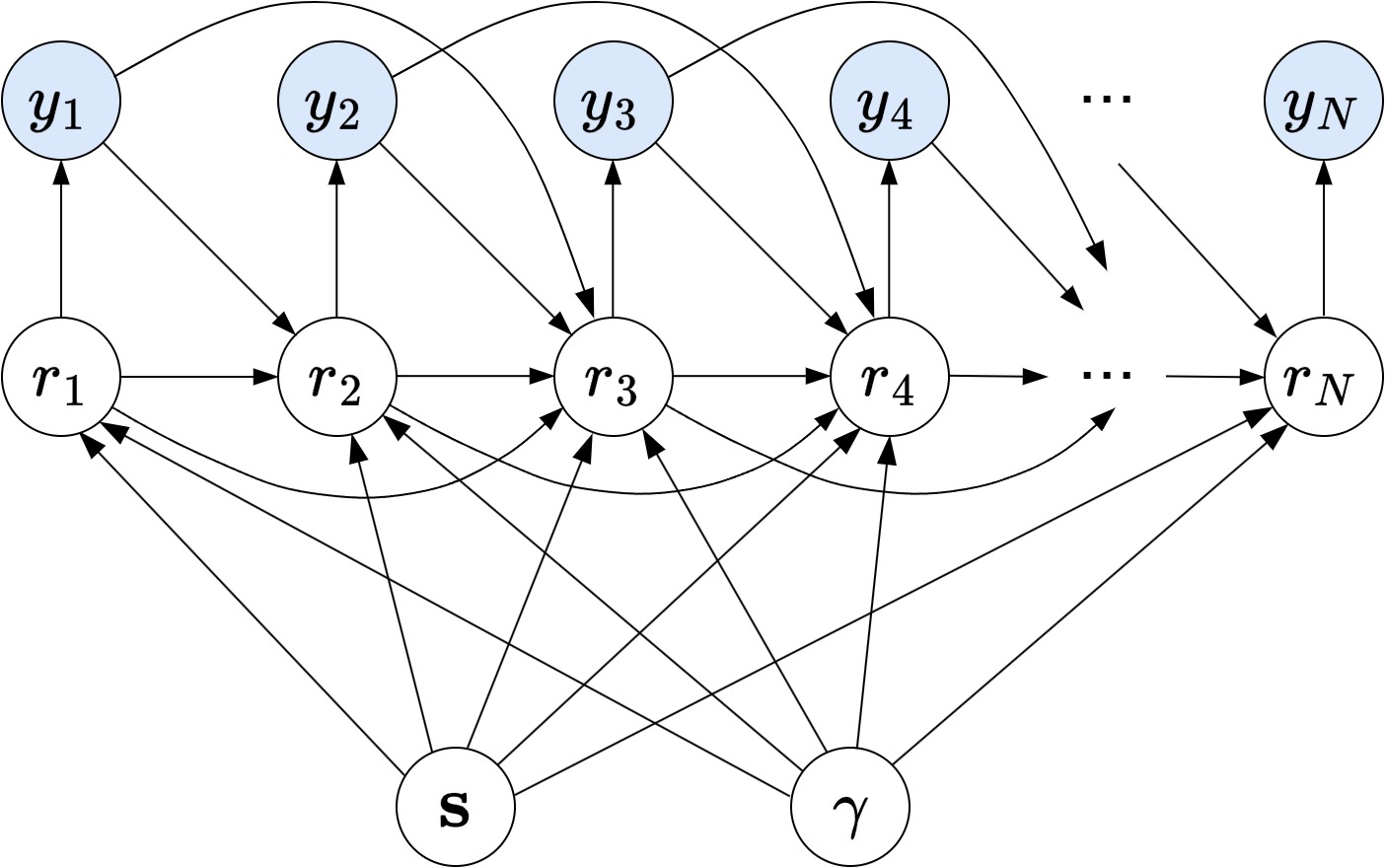}
    \caption{Bayesian network for the $2^{\nd}$-order $\sd$ receiver.}
    \label{fig:Bayesian-Sigma-Delta-2nd}
\end{figure}
Based on \eqref{eq:inout}, the dependency between random variables under spatial $2^{\nd}$-order $\Sigma\Delta$ processing can be illustrated through the graphical Bayesian network model in Fig.~\ref{fig:Bayesian-Sigma-Delta-2nd}. The joint distribution $p(\mb{y},\mb{r},\mb{s},\gamma;\mb{H})$ can be factored as
\begin{align} \label{joint:dist:2nd}
&p(\mb{y},\mb{r},\mb{s},\gamma;\mb{H}) \\
&\quad= p(\mb{s}) p(\gamma)\prod_{i=1}^N p(y_i|r_i) p(r_i|r_{i-1},y_{i-1},r_{i-2},y_{i-2},\mb{s},\gamma;\mb{H}),\nonumber
\end{align}
where the conditional probability 
\begin{align}
    &p(r_i|r_{i-1},y_{i-1}, r_{i-2},y_{i-2}, \mb{s},\gamma;\mb{H}) = \label{eq:42} \\
    &\mathcal{CN}\big(r_i;\mb{h}_i^T\mb{s} + 2\e^{-j\phi}(r_{i-1} \!-\! y_{i-1}) \!-\! \e^{-j2\phi}(r_{i-2} \!-\! y_{i-2}),1/\gamma\big). \nonumber
\end{align}
 {This conditional probability follows a complex Gaussian distribution, driven by the input-output relationship of the $2^{\nd}$-order $\sd$ approach in \eqref{eq:inout}, where the fact that the noise  $n_i$ is Gaussian implies that~\eqref{eq:42} is Gaussian.}
The mean-field VB inference method aims to optimize the variational distribution $q(\mb{s},\mb{r},\gamma)$ such that
\begin{align}
p(\mb{s},\mb{r},\gamma|\mb{y};\mb{H}) \approx q(\mb{s},\mb{r},\gamma) = \prod_{k=1}^K q(s_k) \prod_{i=1}^N q(r_i)q(\gamma).
\end{align}
%
\textit{1) Updating $r_i$:} For $i=1,\ldots,N-2$, the variational distribution $q(r_i)$ can be obtained by taking the expectation of the conditional in \eqref{joint:dist:2nd} w.r.t. $q(\mb{s},\gamma)$ as
\begin{align} \label{q:r_i:2nd}
  \!q(r_i)&\propto \exp\big\{\blr{ \ln p(r_i|r_{i-1},y_{i-1},r_{i-2},y_{i-2},\mb{s},\gamma;\mb{H}) \nonumber\\
  & \hskip2em + \ln p(y_i|r_i) + \ln p(r_{i+1}|r_i,y_i,r_{i-1},y_{i-1},\mb{s},\gamma;\mb{H}) \nonumber\\
  &\hskip2em  + \ln p(r_{i+2}|r_{i+1},y_{i+1},r_{i},y_{i},\mb{s},\gamma;\mb{H})}\big\}, 
\end{align}
which is expanded as in~\eqref{q:r_i:2nd:expand} on the next page,
\begin{figure*}[!th]
\begin{align} \label{q:r_i:2nd:expand}
q(r_i) 
  &\propto p(y_i|r_i) \exp\big\{ - \blr{ \gamma |r_i -  \mb{h}_i^T\mb{s} - 2\e^{-j\phi}(r_{i-1} - y_{i-1}) + \e^{-j2\phi}(r_{i-2} - y_{i-2})|^2} - \blr{ \gamma | r_{i+1} - \mb{h}_{i+1}^T\mb{s}  \nonumber\\
   &\hskip1em - 2\e^{-j\phi}(r_{i} - y_{i}) + \e^{-j2\phi}(r_{i-1} - y_{i-1})|^2}  - \blr{ \gamma |r_{i+2} -\mb{h}_{i+2}^T\mb{s} - 2\e^{-j\phi}(r_{i+1} - y_{i+1})  + \e^{-j2\phi}(r_{i} - y_{i})|^2} \big\}\nonumber\\
   &\propto p(y_i|r_i) \exp\big\{ - \lr{ \gamma}\blr{ |r_i -  \mb{h}_i^T\mb{s} - 2\e^{-j\phi}(r_{i-1} - y_{i-1}) + \e^{-j2\phi}(r_{i-2} - y_{i-2})|^2} - \lr{ \gamma}\blr{ |2(r_{i} - y_{i})- \e^{j\phi}r_{i+1}  \nonumber\\
   &\hskip1em  + \e^{j\phi}\mb{h}_{i+1}^T\mb{s} - \e^{-j\phi}(r_{i-1} - y_{i-1})|^2} - \lr{ \gamma}\blr{ |r_{i} - y_{i}  + \e^{j2\phi}r_{i+2} -\e^{j2\phi}\mb{h}_{i+2}^T\mb{s} - 2\e^{j\phi}(r_{i+1} - y_{i+1})|^2}\big\}\nonumber\\
    &\propto \mathbbm{1}(r_i \in [y_i^\low,y_i^\up]) \exp\big\{\! - \!\lr{ \gamma} (|r_{i} - c_i|^2  + |2r_{i} - d_i|^2 +  |r_i -  f_i|^2)\big\}\nonumber\\
    &\propto \mathbbm{1}(r_i \in [y_i^\low,y_i^\up])\exp\big\{ - 6\lr{ \gamma} \big|r_{i} - (c_i + 2d_i + f_i)/6\big|^2 \big\}\nonumber\\
    &\propto \mathbbm{1}(r_i \in [y_i^\low,y_i^\up])\,\mathcal{CN}\big(r_i;(c_i + 2d_i + f_i)/6,1/(6\lr{ \gamma})\big)
\end{align}
\setlength{\arraycolsep}{1pt} \hrulefill
\setlength{\arraycolsep}{0.0em} \vspace*{1pt}
\end{figure*}
where 
\begin{align}
    &c_i \triangleq \mb{h}_i^T\lr{\mb{s}} \!+\! 2\e^{-j\phi}(\lr{r_{i-1}} \!-\! y_{i-1}) - \e^{-j2\phi}(\lr{r_{i-2}} - y_{i-2}),\nonumber\\
    &d_i \triangleq 2y_{i} + \e^{j\phi}\lr{r_{i+1}} - \e^{j\phi}\mb{h}_{i+1}^T\lr{\mb{s}} \!+\! \e^{-j\phi}(\lr{r_{i-1}} - y_{i-1}),\nonumber\\
    &f_i \triangleq y_{i} - \e^{j2\phi}\lr{r_{i+2}} +\e^{j2\phi}\mb{h}_{i+2}^T\lr{\mb{s}} + 2\e^{j\phi}(\lr{r_{i+1}} - y_{i+1}).\nonumber
\end{align}
Similarly, we can obtain the variational distribution $q(r_{N-1})$ as
\begin{align} \label{q:r_N_1:2nd}
    q(r_{N-1}) &\propto \mathbbm{1}(r_{N-1} \in [y_{N-1}^\low,y_{N-1}^\up]) \nonumber\\
    &\; \times \mathcal{CN}\big(r_{N-1};(c_{N-1} + 2d_{N-1})/5,1/(5\lr{ \gamma})\big),
\end{align}
and $q(r_N)$ as
\begin{align} \label{q:r_N:2nd}
    q(r_{N}) &\propto \mathbbm{1}(r_{N} \in [y_{N}^\low,y_{N}^\up])\,\mathcal{CN}\big(r_{N};c_{N},1/\lr{ \gamma}\big).
\end{align}
%
Defining the residual term $u_i$ as 
\begin{align}
    u_i =& \lr{r_i} - \mb{h}_i^T\lr{\mb{s}} - 2\e^{-j\phi}(\lr{r_{i-1}} - y_{i-1}) \nonumber\\
    &+ \e^{-j2\phi}(\lr{r_{i-2}} - y_{i-2}),
\end{align}
we have $c_i = \lr{r_i} - u_i$, $d_i = 2\lr{r_i} + \e^{j\phi} u_{i+1}$, and $f_i = \lr{r_i} - \e^{j2\phi} u_{i+2}$. We then denote variable $v_i$ as in \eqref{eq:vi}, shown at the top of the next page.
\begin{figure*}[!th]
\begin{align} \label{eq:vi}
v_i &= \left\{\begin{array} {ll}
\frac{c_i+ 2d_i + f_i}{6} = \lr{r_i} - \frac{u_i-2\e^{j\phi}u_{i+1} + \e^{j2\phi}u_{i+2}}{6} & \textrm{ for } i=1,\ldots,N-2, \\
\frac{c_{N-1} + 2d_{N-1}}{5} =  \lr{r_{N-1}} - \frac{u_{N-1}-2\e^{j\phi}u_{N}}{5} & \textrm{ for } i= N-1, \\
c_N =  \lr{r_{N}} - u_{N} & \textrm{ for } i= N. \\
\end{array}
\right.
\end{align}
\setlength{\arraycolsep}{1pt} \hrulefill
\setlength{\arraycolsep}{0.0em} 
\end{figure*}

\textit{2) Updating $s_k$:} The variational distribution $q(s_k)$ is obtained by taking the expectation of the conditional in \eqref{joint:dist:2nd} w.r.t. $q(\mb{r},\gamma)$:
\begin{align}
  q(s_k) &\propto \exp\bigg\{\bblr{ \ln p(s_k) \nonumber\\
  &\hskip2em+ \sum_{i=1}^{N}\ln p(r_{i}|r_{i-1},y_{i-1},r_{i-2},y_{i-2},\mb{s},\gamma;\mb{H}) } \bigg\},\nonumber\\
&\propto p(s_k) \exp\bigg\{-\lr{ \gamma} \sum_{i=1}^{N}\blr{|r_i -  \mb{h}_i^T\mb{s}  \nonumber\\
&\hskip2em - 2\e^{-j\phi}(r_{i-1} - y_{i-1}) + \e^{-j2\phi}(r_{i-2} - y_{i-2})|^2} \bigg\} \nonumber \\
& \propto p(s_k) \prod_{i=1}^N \mathcal{CN} \big(z_{i,k};{h}_{i,k}s_k, \langle \gamma \rangle ^{-1}\big),
\end{align}
where $z_{i,k}$ is defined as 
\begin{align}
    z_{i,k} &= \lr{r_{i}} - 2\e^{-j\phi}(\lr{r_{i-1}} - y_{i-1}) + \e^{-j2\phi}(\lr{r_{i-2}} - y_{i-2}) \nonumber\\
    &\quad - \mb{h}_{i}^T \lr{\mb{s}} + h_{i,k} \langle s_k \rangle \nonumber \\
    &= u_i - h_{i,k}\lr{s_k}.
\end{align}
The variational mean $\lr{s_k}$ and variance $\tau_{s_k}$ then can be obtained accordingly. 

\textit{3) Updating $\gamma$:} 
The variational distribution $q(\gamma)$ can be obtained by taking the expectation of the conditional distribution in \eqref{joint:dist:2nd} w.r.t. $q(\mb{s}, \mb{r})$ as
\begin{align}
q(\gamma) &\propto \exp\bigg\{\Blr{ \ln p(\gamma) \nonumber\\
& \;\;\;+ \sum_{i=1}^N\ln p(r_i|r_{i-1},y_{i-1},r_{i-2},y_{i-2},\mb{s},\gamma;\mb{H}) } \bigg\},
\end{align}
which is expanded as in~\eqref{eq:q:gamma} on the next page,
\begin{figure*}[!th]
\begin{align} \label{eq:q:gamma}
q(\gamma)&\propto \exp \bigg\{(N+\alpha-1)\ln\gamma  -\beta \gamma - \gamma \sum_{i=1}^N \blr{ | r_i -  \mb{h}_i^T\mb{s} - 2\e^{-j\phi}(r_{i-1} - y_{i-1}) + \e^{-j2\phi}(r_{i-2} - y_{i-2})|^2} \bigg\}\nonumber\\
&\propto \exp \bigg\{(N+\alpha-1)\ln\gamma  -\beta \gamma - \gamma \sum_{i=1}^N \big[|u_i|^2 + \tau_{r_i} + \mb{h}_i^H \bs{\Sigma}_{\mb{s}} \mb{h}_i + 4\tau_{r_{i-1}} + \tau_{r_{i-2}}\big]\bigg\}\nonumber\\
&\propto \exp \big\{(N+\alpha-1)\ln\gamma  - \gamma( \beta + \|\mb{u}\|^2 +  6\tr\{\bs{\Sigma}_{\mb{r}}\} + \tr \big\{ \mb{H}\bs{\Sigma}_{\mb{s}}\mb{H}^H\big\} -5\tau_{r_{N}}-\tau_{r_{N-1}}) \big\}, 
\end{align}
\setlength{\arraycolsep}{1pt} \hrulefill
\setlength{\arraycolsep}{0.0em} \vspace*{1pt}
\end{figure*}
where $u_i$ denotes the residual term at receive antenna $i$.
The variational distribution $q(\gamma)$ is Gamma with mean 
\begin{align}
    \langle \gamma \rangle = \frac{N+ \alpha}{\beta\! + \|\mb{u}\|^2\!+ 6 \tr\{\bs{\Sigma}_\mb{r}\}\! - 5\tau_{r_N}\! - \tau_{r_{N-1}}  \!+\! \tr \big\{ \mb{H}\bs{\Sigma}_{\mb{s}}\mb{H}^H\big\}}.
\end{align}


The proposed VB method for MIMO detection with $2^{\nd}$-order $\sd$ quantization is summarized in Algorithm~\ref{algo-2}, where the parameter $\epsilon$ is the numerator of $\hat{\gamma}$ in the variational distributions of \eqref{q:r_i:2nd}, \eqref{q:r_N_1:2nd}, and \eqref{q:r_N:2nd} for updating $r_i$, $r_{N-1}$, and $r_N$, respectively.  {We refer this algorithm as the $2^{\nd}$-order SD-VB algorithm.}

\begin{algorithm}[t]
\small
	\caption{\normalsize -- VB Algorithm for MIMO Detection with $2^{\nd}$-Order $\Sigma\Delta$ Quantization}
	\begin{algorithmic}[1]
		\STATE \textbf{Input:}
		$\mb{y}, \mb{H}$
        \STATE \textbf{Output:}	$\hat{\mb{s}}$
		\STATE Initialize $\hat{r}_i^1 = y_i^1$, ${\tau}_{r_i}^1=0$, $\forall i$, 
  $\hat{s}_k^1=0$, ${\tau}_{s_k}^1= {\rm Var}_{p(s_k)}[s_k]$, $\forall k$, $\mb{u}= \hat{\mb{r}}^1 - \mb{H} \hat{\mb{s}}^1$.
        \FOR{$t=1,2,\ldots$}
        \STATE $\hat{\gamma}^t \leftarrow (N+\alpha)/\big(\beta + \|\mb{u}\|^2 + 6 \tr\{\bs{\Sigma}_\mb{r}\} - 5\tau_{r_N}^t  -\tau_{r_{N-1}}^t $
        \STATE \quad\quad\quad\quad\quad\quad\quad$ + \tr \{\mb{H}\bs{\Sigma}_{\mb{s}}\mb{H}^H\}\big)$
         \FOR{$i=1,\ldots,N$}
         \IF{$i=N$} 
         \STATE $v_N^t \leftarrow \hat{r}_N^t - u_N$, $\epsilon = 1$
         \ELSIF{$i=N-1$}
         \STATE $v_{N-1}^t \leftarrow \hat{r}_{N-1}^t - (u_{N-1} - 2\e^{j\phi}u_{N})/2$, $\epsilon = 5$
         \ELSE 
        \STATE $v_i^t \leftarrow \hat{r}_i^t - (u_i - 2\e^{j\phi}u_{i+1}+\e^{j2\phi}u_{i+2})/6$, $\epsilon = 6$
        \ENDIF
        \STATE $\hat{r}_i^{t+1} \leftarrow \ms{F}_r(v_i^t,\epsilon\hat{\gamma}^t,y_i^{\low},y_i^{\up})$
        \STATE ${\tau}_{r_i}^{t+1} \leftarrow \ms{G}_r(v_i^t,\epsilon\hat{\gamma}^t,y_i^{\low},y_i^{\up})$
        \STATE $u_i \leftarrow u_i - \hat{r}_i^t + \hat{r}_i^{t+1}$
        \STATE $u_{i+1} \leftarrow u_{i+1} + 2\e^{-j\phi} (\hat{r}_i^t - \hat{r}_i^{t+1})$ only for $i<N$
        \STATE $u_{i+2} \leftarrow u_{i+2} - \e^{-j2\phi} (\hat{r}_i^t - \hat{r}_i^{t+1})$ only for $i<N-1$
        \ENDFOR
        \FOR{$k=1,\ldots,K$}
        \STATE $\mb{z}_k^t \leftarrow  \mb{h}_k\hat{s}_k^t+ \mb{u}$
        \STATE $s_k^{t+1} \leftarrow \ms{F}_s(\mb{z}^t_k,\mb{h}_k,\hat{\gamma}^t)$
        \STATE $\tau_{s_k}^{t+1} \leftarrow \ms{G}_{s}(\mb{z}^t_k,\mb{h}_k,\hat{\gamma}^t)$
        \STATE $\mb{u} \leftarrow \mb{u} + \mb{h}_k(\hat{s}_k^t - \hat{s}_k^{t+1})$
        \ENDFOR
        \ENDFOR
        \STATE $\forall k: \hat{s}_k \leftarrow \arg \max_{a\in \mathcal{S}}p_a \mathcal{CN}\big(\mb{z}_{k}^t;\mb{h}_k a,(1/\hat{\gamma}^t)\mb{I}_M\big)$
	\end{algorithmic} 
    \label{algo-2}
\end{algorithm}



{\emph{Remark 3:} The $1^{\rm st}$- and $2^{\rm nd}$-order SD-VB algorithms are designed to address the nonlinear quantization effects of spatial $\sd$ ADCs in massive MIMO systems, particularly for discrete constellations such as PSK and QAM in angularly sectored mmWave channels. The SD-VB algorithms are not optimal for Gaussian input signals because they approximate the posterior of the discrete latent variables and rely on mean-field factorization, which does not fully capture the continuous statistical properties of Gaussian signals. 
}

The $1^{\rm st}$- and $2^{\rm nd}$-order SD-VB algorithms, based on CAVI, factorize $r_i$, $s_k$, and $\gamma$, iteratively updating $q(s_k)$, $q(r_i)$, and $q(\gamma)$ until they converge to a locally optimal solution. This efficient approach suits massive MIMO systems with highly non-linear SD quantization, where local optima yield acceptable SER. We will show through simulations in Section V that these algorithms achieve low SER, outperforming state-of-the-art methods, especially with optimized antenna spacing, with the $2^{\rm nd}$-order SD-VB enhancing detection by reaching a better local optimum under various scenarios and mutual coupling effects.

\subsection{Computational Complexity Analysis}

\begin{table}[t]
\begin{tabular}{|c | c|} 
 \hline
 \bf{Algorithm} & \bf{Complexity} \\ [0.5ex] 
 \hline\hline
MF-VB in \cite{nguyen2022variational} & $\mathcal{O}(NKT + |\mathcal{S}|KT)$ \\ \hline
MF-QVB in \cite{nguyen2023variational} & $\mathcal{O}(NKT + |\mathcal{S}|KT)$  \\ \hline
LMMSE & $\mathcal{O}(N^3 + |\mathcal{S}|K)$ \\ \hline
$1^{\st}$-order SD-VB & $\mathcal{O}(NKT + |\mathcal{S}|KT)$  \\ \hline
$2^{\rm nd}$-order SD-VB  & $\mathcal{O}(NKT + |\mathcal{S}|KT)$  \\ \hline
\end{tabular}
\caption{Algorithm Computational Complexity}
\label{table}
\end{table}

The computational complexity of the proposed algorithms is analyzed here assuming $N \ge K$. For iterative algorithms, the complexity order is given per iteration. For LMMSE, the computation of $\mb{H}\bs{\Sigma}_\mb{s}\mb{H}^H$ for the linear detector in~\eqref{LMMSE:detector} requires a complexity of $\mathcal{O}(NK^2)$, while calculating $\mb{U}^{-1}\bs{\Sigma}_{\mb{q}} \mb{U}^{-H}$ requires $\mathcal{O}(N^3)$ operations due to the matrix inversion. In addition, the denoiser for $K$ users contributes a complexity of $\mathcal{O}(|\mathcal{S}|K)$. Thus, the overall computational burden of the LMMSE detector is dominated by the matrix inversion, resulting in an overall complexity of order $\mathcal{O}(N^3 + |\mathcal{S}|K)$. For the $1^{\st}$- and $2^{\nd}$-order SD-VB algorithms in Algs. $1$ and $2$, respectively, the cost to compute $\tr \{\mb{H}\bs{\Sigma}_{\mb{s}}\mb{H}^H\}= \sum_{k=1}^K \tau_{s_k}\|\bar{\h}_k\|^2$ in step~5 is $\mathcal{O}(NK)$, which dominates that for computing the residual vector $\|\mb{u}\|^2$ which is $\mathcal{O}(N)$. The computation in step~18 of Algorithm~\ref{algo-1} and step~22 of Algorithm~\ref{algo-2} adds a complexity of $\mathcal{O}(|\mathcal{S}|)$ for each user. The overall per-iteration complexity of Algs. $1$ and $2$ is thus $\mathcal{O}(NK + |\mathcal{S}|K)$, and we see that the $1^{\st}$- and $2^{\nd}$-order SD-VB algorithms have the same complexity as that for the MF-VB and MF-QVB approaches, although the $\sd$ methods can provide a significant improvement in MIMO detection performance under the given assumptions of spatial oversampling or sectored users.  {Note that the complexity of the VB-based algorithms listed in Table I also reflects the number of iterations $T$ needed for convergence. Numerical simulations suggest that $T$ is typically below $20$.}

\section{Numerical Results} \label{sec:Numerical}
In this section, we present illustrative numerical results for the performance of the proposed $1^{\st}$- and $2^{\nd}$-order SD-VB algorithms compared with state-of-the-art data detection methods such as the matched-filter quantized VB (MF-QVB) in \cite{nguyen2023variational} and LMMSE-based detection for various scenarios. 
We implement all VB-based algorithms with a maximum of $50$ iterations and consider scenarios with $100$ transmitted data symbols.
 {The channel matrix 
$\mb{H}$ is normalized such that each column has unit norm, i.e., $\mathbb{E}\big[\|\bar{\mb{h}}_k\|^2\big] = 1, \forall k$ \cite{nguyen2022variational}. The SNR is defined as
\begin{align}
    {\rm SNR} = \frac{\mathbb{E}\big[\|\mb{Hs}\|^2\big]}{\mathbb{E}\big[\|\mb{n}_c\|^2\big]} =\frac{K}{ \tr\{\mb{R}\}}.
\end{align}
If the system is without mutual coupling, the noise covariance matrix $\mb{R}$ reduces to $N_0 \mb{I}$, and the SNR becomes ${{\rm SNR} = K/(N N_0)}$}.
Unless otherwise stated, all cases assume the number of paths as $L=20$, the width of the angular sector as $\Theta = 40^\circ$ and assume it is centered at $\theta_0 = 0^\circ$. We assume all users lie within the same azimuth angular range, with AoAs drawn uniformly from the interval $[-20^\circ,20^\circ]$. 
We also assume the number of BS antennas is $N=128$, the number of users is $K=16$, the phase shift of the $\sd$ array is $\phi = 2\pi \frac{d}{\lambda}\sin(\theta_0)$, $d = \lambda/6$, and the users employ QPSK signaling. 
 {Taking the mutual coupling effect into account, we set the circuit parameters in~\eqref{eq:mc_mat} and \eqref{eq:mutual_coupling} as $\sigma_i^2=2kTB/R$, and $\sigma_u^2=2kTBR$, leading to the noise ratio $R_N =\sqrt{\sigma_u^2/\sigma_i^2}
=R$ where $R=50~\Omega$, $T=290~\mathrm{K}$, $\rho=0$, and $B=20~\mathrm{MHz}$. The factor of 2 in the relationship $\sigma_n^2 = 2kTBR$ appears because we are accounting for noise in both the antennas and the LNAs.}

To highlight the effectiveness of the proposed approach, we compare the performance of the SD-VB algorithms with that of the following benchmark approaches:
\begin{enumerate}
    \item The LMMSE receiver presented in Section~\ref{sec:LMMSE} implemented with the $1^{\st}$-order $\sd$ architecture and $1$-bit quantizers, 
    \item The MF-QVB algorithm developed in \cite{nguyen2023variational} implemented with conventional few-bit quantizers, 
    \item The MF-VB algorithm developed in \cite{nguyen2022variational} implemented with ideal/infinite quantizers.
\end{enumerate}

\begin{figure}[!t]
	\centering
	\includegraphics[width=1.05\linewidth]{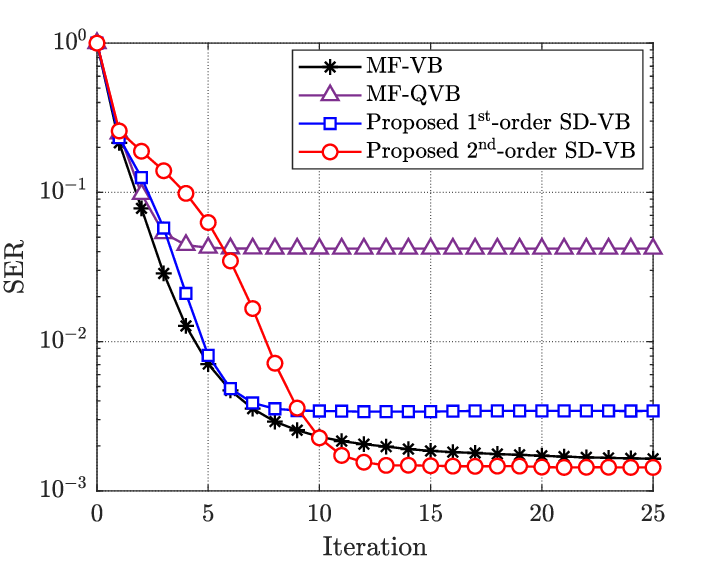}
	\caption{ {Convergence of the iterative algorithms with SNR$=12$ dB, $\theta_0 = 20^\circ$, $d = \lambda/6$, $\Theta = 40^\circ$, $3$-bit quantization, and QPSK signaling. The plots are obtained by averaging over 200 trials.}}
	\label{Fig_SER_Con}
\end{figure}

{ {We first study the convergence of all iterative algorithms with $3$-bit quantization, as shown in Fig.~\ref{Fig_SER_Con}. All algorithms are observed to converge in less than 20 iterations, which demonstrates the effectiveness of using CAVI in our VB framework.}}

\begin{figure}[!t]
	\centering
\includegraphics[width=1.05\linewidth]{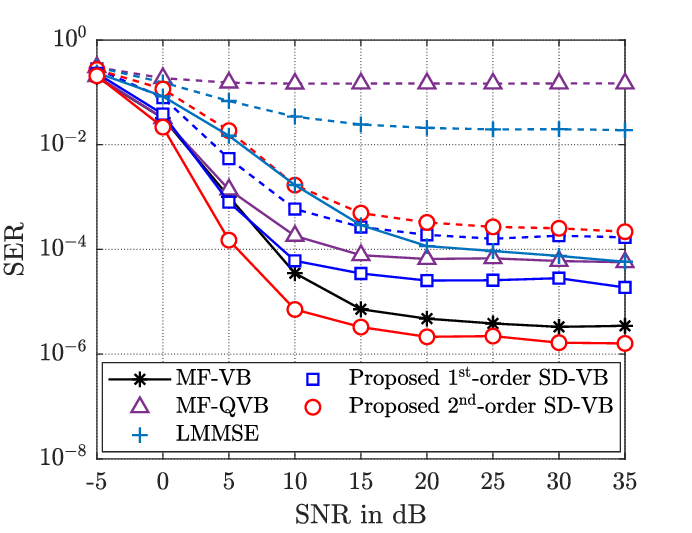}
	\caption{ {SER performance vs. SNR with $\theta_0 = 20^\circ$, $\Theta = 40^\circ$, $d = \lambda/4$, QPSK, and without mutual coupling. For the quantized system, solid lines represent results with 3-bit ADCs; dashed lines represent results with 1-bit ADCs.}}
	\label{Fig_SER_SNR_QPSK_1_3_bit}
\end{figure}

\begin{figure}[!t]
	\centering
	\includegraphics[width=1.09\linewidth]{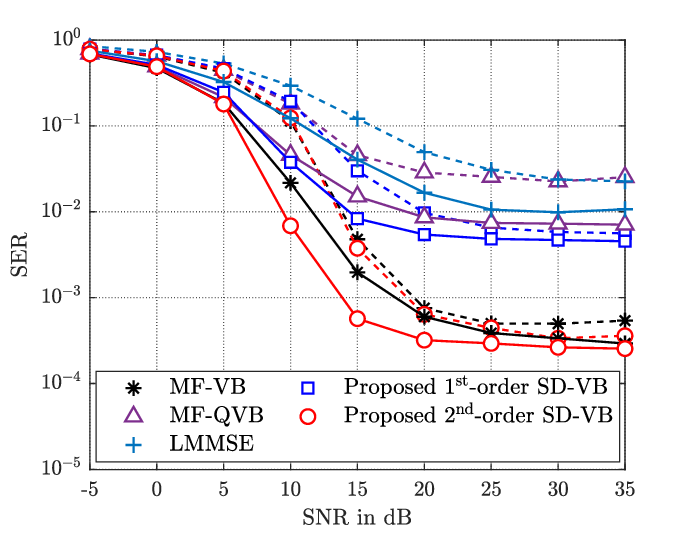}
	\caption{ {SER performance vs. SNR with 3-bit quantizer, $\theta_0 = 20^\circ$, $\Theta = 40^\circ$, $d = \lambda/4$, $16$-QAM. Solid lines represent results without mutual coupling; dashed lines represent results with mutual coupling effects.}}
	\label{Fig_SER_SNR_16QAM_MC}
\end{figure}

 {We show the data detection performance improvement of the proposed SD-VB algorithms using 1-bit and 3-bit ADCs with QPSK in Fig.~\ref{Fig_SER_SNR_QPSK_1_3_bit}, and using 3-bit ADCs with $16$-QAM and MC effects in Fig. \ref{Fig_SER_SNR_16QAM_MC}. The SD-VB algorithms outperform MF-QVB for both 1-bit and 3-bit ADCs} since the $\sd$ receivers shape the quantization noise away from the spatial frequencies of interest. The SD-VB algorithms also significantly outperform the LMMSE-based receiver because the latter's reliance on linear processing makes it ineffective at separating users with highly correlated channels.  {The $1^{\st}$- and $2^{\nd}$-order SD-VB algorithms with $1$-bit quantization and QPSK exhibit higher SER than MF-VB since the latter operates as an unquantized (full-precision) system.
We note that coarse quantization (e.g., 1-bit or 3-bit ADCs) introduces quantization noise that cannot be reduced and persists even at high SNR. As a result, the symbol error rates (SERs) of all algorithms processing quantized signals converge to their error floors at high SNR, since quantization noise becomes the dominant source of error.}

The performance of $2^{\nd}$-order SD-VB significantly improves when the ADC resolution is increased from $1$ to $3$ bits, achieving the best SER among the considered algorithms, while MF-QVB performs the worst. This advantage is due to the more aggressive noise shaping of the $2^{\nd}$-order $\sd$ architecture, greatly reducing the quantization noise in lower spatial frequency ranges.
 {Although MF-QVB shows some improvement with $3$-bit quantization for both QPSK and 16-QAM}, it remains inefficient for cases with users confined to an angular sector since the quantization noise is uniformly spread across all spatial frequencies. While the MF-VB algorithm employs ideal/infinite quantization, it also experiences an error floor due to the channel correlation \cite{nguyen2022variational}. The $2^{\nd}$-order SD-VB with only $3$-bit quantizers outperforms MF-VB since it implicitly exploits knowledge of the angular sector to which the users are confined and the resulting spatial correlation.
While the LMMSE receiver provided in \eqref{LMMSE:detector} is designed for $1$-bit quantization, it shows considerable SER improvement when the quantization resolution is increased to $3$ bits. Thus, in subsequent simulations we will only show results for the $3$-bit LMMSE implementation. 

 {The effect of MC is shown in Fig.~\ref{Fig_SER_SNR_16QAM_MC} for 16-QAM signaling, illustrating that the SER performance of LMMSE and MF-QVB degrades considerably with MC, while the proposed SD-VB algorithms have only a slight decrease at high SNRs. This is due to two reasons. First, since spatial $\sd$ techniques rely on the correlation between signals at adjacent antennas, the correlation introduced by MC actually provides some benefit for $\sd$ that counteracts the corresponding loss due to channel correlation. In addition, when MC introduces colored noise, the $\sd$ noise-shaping effect treats it similarly to quantization noise, pushing it outside the signal bandwidth.
}
{The LMMSE detector is implemented using \eqref{LMMSE:detector:MC}, which incorporates the known colored noise covariance matrix $\mb{R}$ induced by MC.
Despite this, the LMMSE approach still exhibits SER degradation compared to SD-VB. This is due to the limitations of the Bussgang decomposition, which assumes Gaussian inputs, and the approximation of the quantization noise covariance $\bs{\Sigma}_{\mb{q}}^{\rm{MC}}$ as diagonal \cite{pirzadeh2020effect,pirzadehmc}.}
%

\begin{figure}[!t]
	\centering
	\includegraphics[width=1.0\linewidth]{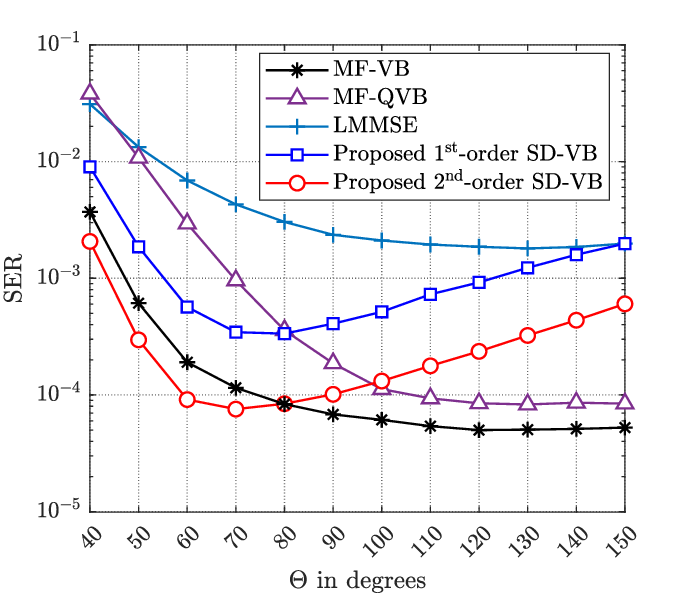}
	\caption{SER performance vs. azimuth angular spread $\Theta$, with 3-bit quantizers, SNR $ = 5$ dB, $\theta_0 = 0^\circ$, ${d = \lambda/6}$, and QPSK.}
	\label{Fig_SER_Theta}
\end{figure}

Fig.~\ref{Fig_SER_Theta} shows the effect of the azimuth angular spread on SER performance. For narrow azimuth ranges, all approaches experience high SER due to the extreme channel correlation that results in $K=16$ users densely packed together. The $3$-bit SD-VB implementation has the best performance among all the approaches for sectors smaller than $80^\circ$. Both SD-VB implementations achieve their best performance for $\Theta \in [60^\circ,80^\circ]$, but their SERs degrade as the sectors become wider since the noise-shaping effect is limited. MF-VB and MV-QVB have the best performance for large sectors since the $\sd$-based approaches have reduced spatial correlation to exploit. The advantage of using VB, in general, is evident in the superior performance of the VB algorithms compared with LMMSE in all cases.

\begin{figure}[!t]
	\centering
	\includegraphics[width=1.0\linewidth]{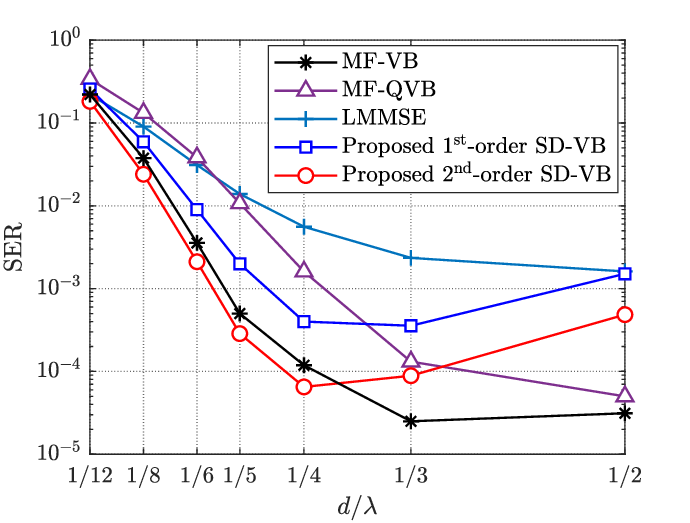}
	\caption{SER performance vs. $d/\lambda$, with 3-bit quantizers, $N = 128$, $\Theta = 40^\circ$, $\theta_0 = 0^\circ$, QPSK, ${\rm SNR} = 5$ dB, and without MC.}
	\label{Fig_SER_Wavelength}
\end{figure}

Fig.~\ref{Fig_SER_Wavelength} presents the effect of antenna spacing and wavelength on the SER performance of all detection algorithms for an array with a fixed number of antennas ($N=128$) without considering MC. Observations similar to those in the previous example can be made here. In particular, when $d$ is very small, the array aperture is significantly reduced and none of the methods are able to counteract the extreme channel correlation that results from the narrow angular sector of $\Theta=40^\circ$. The SER of all algorithms improves with increasing $d$, although the $1^{\st}$- and $2^{\nd}$-order SD-VB algorithms provide the best performance for values of $d$ around $1/4$ to $1/3$, and degrade for larger $d$ since the benefit of oversampling is lost. As the antenna spacing increases, the reduced channel correlation benefits MF-VB and MF-QVB. {Although reducing the antenna spacing to around $1/3$ to $1/4$ wavelength can enhance the SER for the SD-VB algorithms, it reduces the achievable data rate due to the reduced array aperture \cite{pirzadeh2020effect,rao2021massive}. Similarly, lower quantization resolutions (e.g., 1-bit or 3-bit) improve power efficiency but limit the amount of information per sample. As studied in \cite{rao2021massive}, spectral efficiency in $\sd$ MIMO systems scales with the oversampling rate, highlighting a trade-off between noise reduction and capacity. In future works, we will analyze the capacity of MIMO systems with $2^{\rm nd}$-order $\sd$ ADCs, leveraging the probabilistic modeling capabilities of the VB framework. 
}


\begin{figure}[!th]
	\centering
	\includegraphics[width=1.0\linewidth]{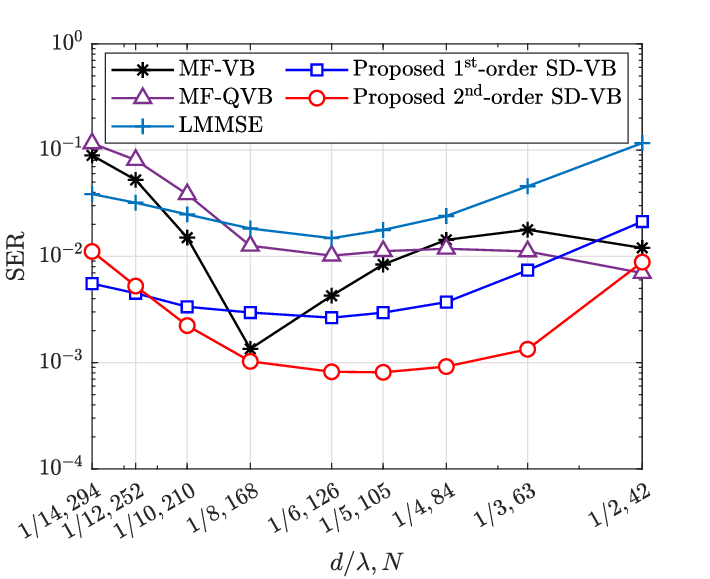}
	\caption{  {SER performance vs. ($d/\lambda$, $N$), with mutual coupling effects, a fixed array $d_0 = 14\,$cm, 3-bit quantizers, QPSK, $K=16$, $\Theta = 40^\circ$, $\theta_0 = 0^\circ$, and ${\rm SNR} = 18$ dB}.}
	\label{Fig_SER_Oversampling_MC}
\end{figure}

In Fig.~\ref{Fig_SER_Oversampling_MC} we study the impact of the antenna spacing $d$ in the presence of MC, but in this case, we fix the array aperture $d_0 = Nd$ so that as $d$ decreases, the number of antennas $N$ increases. For example, if the aperture is set at $d_0 = Nd =14$ cm, the wavelength is $\lambda = c/f = 6.7$ mm at $f = 45$ GHz, and as $d$ decreases from $\lambda/2$ to $\lambda/14$, $N$ increases from $42$ to $294$. As shown in the figure, in this case there is much less degradation for small $d$ since the larger array aperture maintains a more constant level of channel correlation. 
{The SD-VB algorithms maintain superior SER performance over the other algorithms under MC due to their ability to exploit the MC-induced spatial correlation, and because their noise-shaping process treats colored noise introduced by MC similarly to quantization noise, pushing it away from the spatial frequencies of interest.
In this case, the best trade-off between the number of observations and the channel correlation for the SD-VB algorithms occurs for $d=\lambda/6$. LMMSE struggles with high channel correlation under MC since its linear processing cannot effectively de-correlate the users, leading to higher SER. The MF-VB benefits from infinite precision but is still limited by channel correlation, performing worse than the SD-VB algorithms when $d>\lambda/6$ and for narrow sectors.}
{The spatial $\sd$ architecture achieves its best performance with small antenna spacings and confined angular sectors but this potentially limits its real-world application. Nonetheless, these constraints enable compact, power-efficient array designs, e.g., an aperture of 14 cm can fit $126$ elements at $d = \lambda/6$.
}

\begin{figure}[!t]
	\centering
	\includegraphics[width=0.97\linewidth]{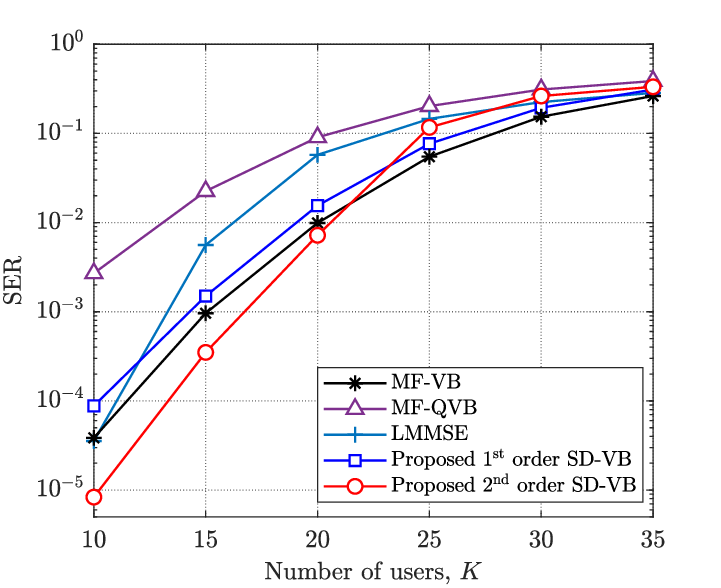}
	\caption{SER performance vs. number of users with 3-bit quantizers, ${\rm SNR} = 8$ dB, QPSK, $N=80$, $d = \lambda/6$, $\theta_0 = 0^\circ$, and $\Theta = 40^\circ$.}
	\label{Fig_SER_user}
\end{figure}

In Fig.~\ref{Fig_SER_user}, we show the effects of the number of users $K$ on the SER performance. When $K$ decreases, the lower spatial channel correlation enhances the data detection performance of all algorithms. The $2^{\nd}$-order SD-VB algorithm achieves the lowest SER among the considered algorithms when $K \le 20$, including MF-VB with infinite precision ADCs. On the other hand, MF-VB has the best performance for large $K$, but in these cases the SER is already quite high. 
MF-QVB performs the worst due to its use of matched filtering to handle multiuser interference, which is suboptimal with high channel correlation, even with relatively few users. LMMSE excels with a small number of users since it is able to more effectively eliminate inter-user interference, achieving a lower SER than MF-VB, $1^{\st}$-order SD-VB, and MF-QVB when $K = 10$.

\begin{figure}[!t]
	\centering
	\includegraphics[width=1.0\linewidth]{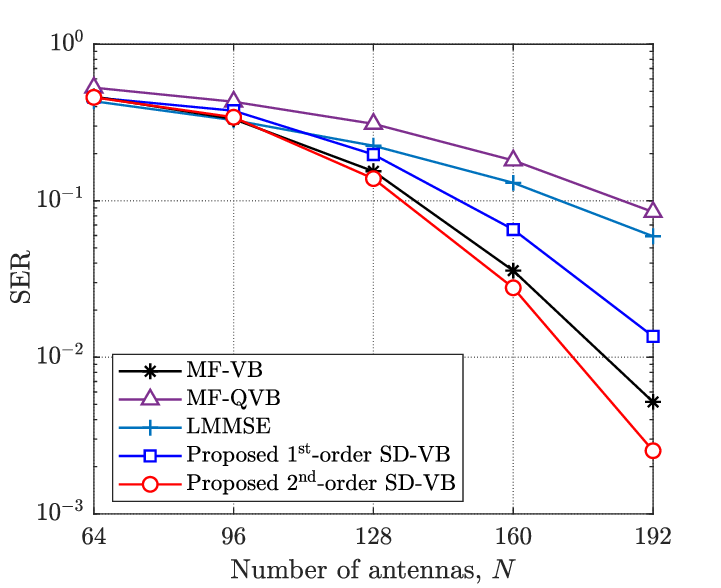}
	\caption{SER performance vs. number of antennas with 3-bit quantizers, SNR$=8$ dB, QPSK, $K=30$, $d = \lambda/6$, $\theta_0 = 0^\circ$, and $\Theta = 40^\circ$.}
	\label{Fig_SER_antenna}
\end{figure}

In Fig.~\ref{Fig_SER_antenna} we present results for data detection with different numbers of antennas. As $N$ increases, the $1^{\st}$- and $2^{\nd}$-order SD-VB algorithms show significant SER improvement, while the improvement is marginal for LMMSE and MF-QVB. The SER performance gap between $2^{\nd}$-order SD-VB and the other detection algorithms becomes wider as the number of antennas increases. 
{Note that MF-VB outperforms $1^{\rm st}$-order SD-VB since it is implemented under the assumption of zero quantization noise. 
}

\begin{figure}[!t]
	\centering
	\includegraphics[width=1.0\linewidth]{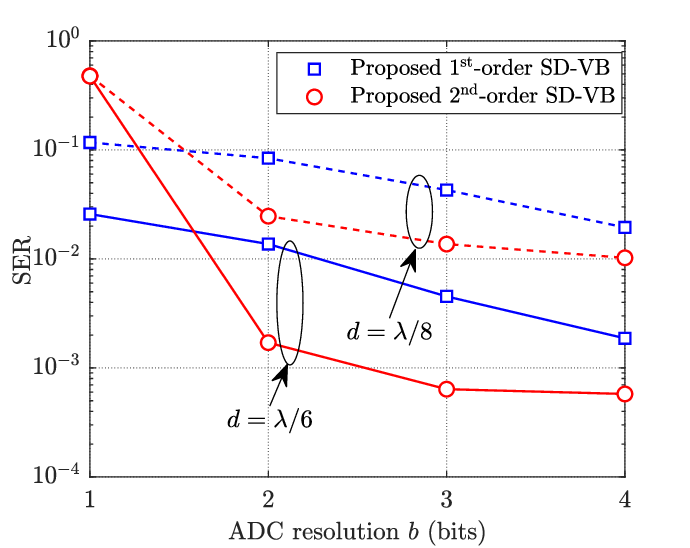}
	\caption{SER performance vs. $\sd$ ADC resolution $b$ bits with SNR$=12$ dB, QPSK, $\theta_0 = 0$, $d = \lambda/6$, and $\Theta = 40^\circ$.}
	\label{Fig_SER_resolution}
\end{figure}

Fig.~\ref{Fig_SER_resolution} displays the SER for the $\sd$ algorithms as a function of the ADC resolution $b$ for $d = \lambda/8$ and $d=\lambda/6$. While the SER decreases as $b$ increases to $3$, there is relatively little improvement for higher ADC resolutions.  
The $1^{\st}$-order SD-VB approach offers better data detection performance than $2^{\nd}$-order SD-VB for $1$-bit quantization. This can occur when the $1$-bit ADCS in a $2^{\nd}$-order $\sd$ architecture becomes overloaded, a condition that occurs when where the input signals combined with the quantization noise exceed the full-scale range of the $1$-bit quantizer leading to instability \cite{aziz1996overview}.
\begin{figure*}[t]
	\subfloat[$\theta_0 = 0^\circ$]{
		\begin{minipage}[c][1\width]{0.325\textwidth}
			\centering
			\includegraphics[width=1.0\textwidth]{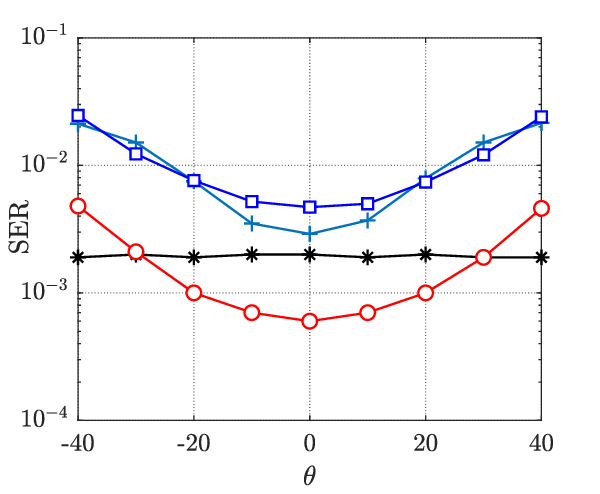}\label{Fig_SER_theta_0}
	\end{minipage}}
	\hfill 	
	\subfloat[$\theta_0 = 30^\circ$]{
		\begin{minipage}[c][1\width]{0.325\textwidth}
			\centering
			\includegraphics[width=1.0\textwidth]{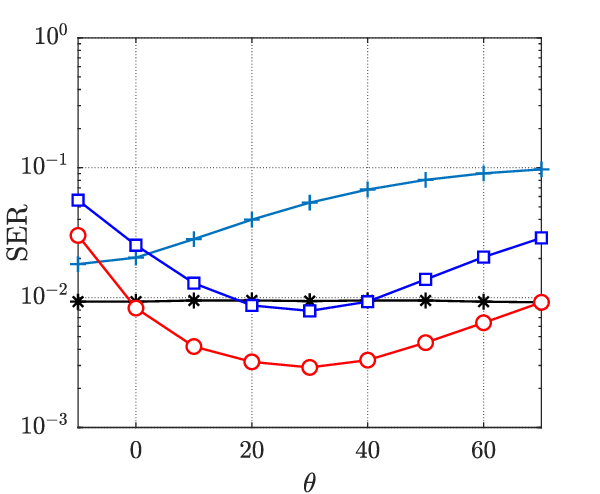}\label{Fig_SER_theta_30}
	\end{minipage}}
	\hfill	
	\subfloat[$\theta_0 = 60^\circ$]{
		\begin{minipage}[c][1\width]{0.325\textwidth}
			\centering
			\includegraphics[width=1.0\textwidth]{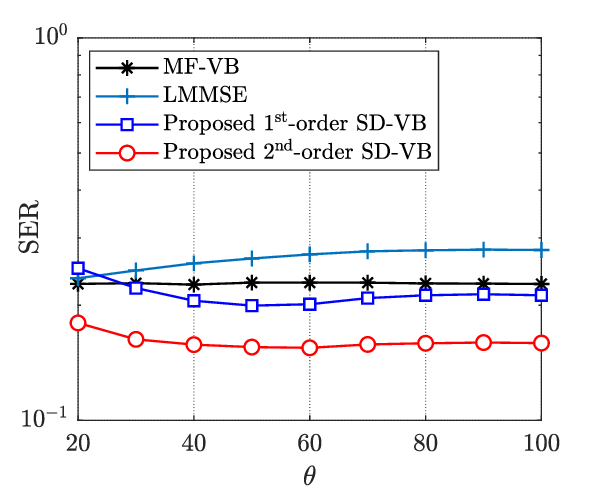}\label{Fig_SER_theta_60}
	\end{minipage}}
	\caption{SER performance vs. steering angles $\phi = 2\pi\frac{d}{\lambda}\sin(\theta)$ of $\sd$ ADCs with 3-bit quantizers, SNR at $12$\,dB, QPSK, and $d = \lambda/6$.} \label{Fig_SER_theta}
\end{figure*}

Fig.~\ref{Fig_SER_theta} shows the SER of the LMMSE, $1^{\st}$-order SD-VB, and $2^{\nd}$-order SD-VB algorithms for $3$-bit $\sd$ architectures implemented with different steering angles $\phi = 2\pi\frac{d}{\lambda}\sin(\theta)$. The steering angles are varied while the center angle of the sector is set as $\theta_0 = 0^\circ, 30^\circ$, and $60^\circ$ in Figs. \ref{Fig_SER_theta}(a), \ref{Fig_SER_theta}(b), and \ref{Fig_SER_theta}(c), respectively. Results for MF-VB are included as a reference, although its performance does not depend on $\phi$ since it does not use the $\sd$ implementation. As expected, the lowest SER for the $1^{\st}$- and $2^{\nd}$-order SD-VB architectures is obtained when the steering angle is chosen to match the center angle of the sector, i.e., $\theta = \theta_0$. 
The SER performance of all algorithms degrades as the center angle of the sector increases, since the ability of the array to spatial separate the signals decreases.
LMMSE performs best without a phase offset, and degrades significantly as the angular sector of the users moves away from broadside. 

\section{Conclusions} \label{sec:Conclusions}
In this paper, we have developed two MIMO detection algorithms based on the VB inference framework in massive MIMO systems with few-bit $\sd$ ADCs. We first modeled the Bayesian networks for the $1^{\st}$- and $2^{\nd}$-order $\sd$ receivers, based on which the variational distributions of the latent variables were obtained in closed form. We then proposed two iterative algorithms for MIMO detection with efficient updates and low-complexity implementation.
Simulation results showed that the proposed $1^{\st}$- and $2^{\nd}$-order SD-VB algorithms achieve their best data detection performance when the users are confined to angular sectors less than $80^\circ$ wide and with antenna spacings on the order of $1/4$ to $1/6$ a wavelength. Under these conditions, a $2^{\nd}$-order SD-VB approach with $3$-bit ADCs can outperform the MF-VB algorithm implemented with infinite resolution quantization, since SD-VB directly exploits the users' spatial correlation.  {The $2^{\nd}$-order SD-VB maintained superior SER performance under MC due to the fact it exploits correlation between signals at adjacent antennas for noise shaping and hence partially offsets the negative effects of channel correlation.}
The simulation results also suggest there are benefits to using $1^{\st}$-order SD-VB rather than the $2^{\nd}$-order SD-VB implementation with $1$-bit quantization due to potential ADC overloading and unstable performance.  {
In future work, we will explore joint grid-less channel estimation and data detection using VB-based techniques for massive MIMO systems with few-bit $\sd$ ADCs.}

\appendices 

\vspace{-0.3cm}
{ 
\section{Linear Model for the $1^{\rm st}$-Order $\sd$ Architecture}\label{appendix-0}
From \eqref{eq:quan:signal}, the input to the quantizers of the $1^{\rm st}$-order $\sd$ architecture is expressed as
\begin{align} \label{eq:r:1st}
    \mb{r} = \mb{U x} -\mb{Vy},
\end{align}
where $\mb{V} = \mb{U} - \mb{I}_N$ and the matrix $\mb{U}$ is given by
\begin{align} \label{eq:matrix:U}
\mb{U} = \begin{bmatrix}
                1 & 0 & \ldots & 0 & 0\\
                \e^{-j\phi} & 1 & \ldots & 0 & 0\\
                \e^{-j2\phi} & \e^{-j\phi} & \ldots & 0  & 0\\
                \vdots & \vdots & \vdots &\ddots & \vdots   \\
                \e^{-j(N-1)\phi} & \e^{-j(N-2)\phi} &  \ldots & \e^{-j\phi} & 1
            \end{bmatrix}.
\end{align}
We model the $1^{\rm st}$-order $\sd$ quantization operation as
\begin{align} \label{eq:y:LMMSE}
    \mb{y} = \mc{Q}(\mb{r}) = \bs{\Gamma} \mb{r} + \mb{q},
\end{align}
where $\bs{\Gamma}$ is an $N\times N$ matrix, and $\mb{q}=\mb{y}-\bs{\Gamma}\mb{r}$ represents the effective quantization noise corresponding to the choice of $\bs{\Gamma}$. Following the approach in \cite{rao2021massive,pirzadeh2020spectral}, we design $\bs{\Gamma}$ using the Bussgang decomposition such that $\mathbb{E}[r_i q_i^*] = 0$ for the corresponding pairs of elements in $\mb{r}$ and $\mb{q}$. This results in a diagonal $\bs{\Gamma}$ whose $i$-th diagonal element $\xi_i$ given by 
$\xi_i = {\mathbb{E}[r_i y_i^*]}/{\mathbb{E}[|r_i|^2]}$.
Substituting \eqref{eq:r:1st} into \eqref{eq:y:LMMSE}, we obtain 
\begin{align} \label{eq:y:LMMSE1}
    \mb{y} &= \mb{\Gamma}\mb{U x} - \mb{\Gamma}\mb{Vy} + \mb{q}, \nonumber\\
   \Leftrightarrow \mb{y}  &= (\mb{I} + \bs{\Gamma} \mb{V})^{-1} \bs{\Gamma} \mb{U x} + (\mb{I} + \bs{\Gamma} \mb{V})^{-1}\mb{q}.
\end{align}
Based on \cite{pirzadeh2020spectral,rao2021massive}, the quantization step size can be chosen such that $\xi_i =1$, $\forall i$, implying $\bs{\Gamma} = \mb{I}$. Using the identity $\mb{U} = \mb{V} + \mb{I}$, \eqref{eq:y:LMMSE1} can be simplified as
\begin{align} \label{eq:y:LMMSE2}
    \mb{y} = (\mb{I} + \mb{V})^{-1} \mb{U x} + (\mb{I} + \mb{V})^{-1}\mb{q} = \mb{x} + \mb{U}^{-1}\mb{q}.
\end{align}
}

\vspace{-0.3cm}
\section{Calculation of $\ms{F}_r (\mu,\gamma,a,b)$ and ${\sf G}_r (\mu,\gamma,a,b)$}\label{appendix-1}
Define $\alpha = \sqrt{2\gamma}(a-\mu)$ and $\beta = \sqrt{2\gamma}(b-\mu)$. For an arbitrary complex random variable $\mathcal{CN}(\mu, \gamma^{-1})$ whose real and imaginary parts are both truncated on the interval $(a,b)$, the mean $\ms{F}_r (\mu,\gamma,a,b)$ and variance ${\sf G}_r (\mu,\gamma,a,b)$ are given by \cite{nguyen2023variational}
\begin{align}
    &\ms{F}_r (\mu,\gamma,a,b) = \mu - \frac{1}{\sqrt{2 \gamma}}\frac{f(\beta) - f(\alpha)}{F(\beta) - F(\alpha)},\\
    &\ms{G}_r (\mu,\gamma,a,b) \!= \!\frac{1}{2{\gamma }}\bigg[1 \!- \!\frac{\beta f(\beta) \!- \!\alpha f(\alpha)}{F(\beta) \!- F(\alpha)} \!-\! \bigg(\!\frac{f(\beta) - f(\alpha)}{F(\beta) - F(\alpha)}\! \bigg)^2 \bigg],
\end{align}
 {where $F(x)$ and $f(x)$ are respectively the cumulative distribution function and probability density function of a logistic random variable $x$ with zero mean and unit variance, defined as $F(x) = 1/(1+\e^{-3x/\sqrt{\pi}})$ and $p(x) = \frac{3}{\sqrt{\pi}}F(x)(1 - F(x))$.}

For the case of 1-bit quantizers, the three quantization thresholds are: $\{-\infty, 0, \infty\}$, simplifying the computation of $\ms{F}_r (\mu,\gamma,a,b)$ and $\ms{G}_r (\mu,\gamma,a,b)$. 
Denoting $\zeta = \mr{sign}(y)\sqrt{2\gamma}\mu$ with quantization level $y$, we have
   \begin{align*}
    &\ms{F}_r (\mu,\gamma,a,b) = \mu + \frac{\mr{sign}(y)}{\sqrt{2 \gamma}}\frac{f(\zeta)}{F(\zeta)},\\
    &\ms{G}_r (\mu,\gamma,a,b) = \frac{1}{2 \gamma }\Bigg[1 - \frac{\zeta f(\zeta)}{F(\zeta)} - \bigg(\frac{f(\zeta)}{ F(\zeta)} \bigg)^2 \Bigg].
\end{align*}
\section{Calculation of ${\sf F}_s(\mb{z}_k,\mb{h}_k,\gamma)$ and ${\sf G}_s(\mb{z}_k,\mb{h}_k,\gamma)$}\label{appendix-2}

Given $\mb{z}_k$ 
in \eqref{eq:zk:SIMO}, the posterior distribution of $s_k$ given $\mb{z}_k$ is 
\begin{align} \label{postDist}
    p(s_k|\mb{z}_k;\mb{h}_k,\gamma) \propto p(s_k) \prod_{i=1}^N \mathcal{CN} \big(z_{i,k};{h}_{i,k}s_k,  \gamma^{-1}\big).
\end{align}
For $a \in \mathcal{S}$, we have
\begin{align} 
    p(s_k \!=\! a|\mb{z}_k;\mb{h}_k,\gamma) =\frac{p_a}{Z} \exp\bigg\{\!\!-\!\gamma \sum_{i=1}^N|z_{i,k}\!-\! h_{i,k} a|^2 \bigg\} ,
\end{align}
where $Z = \sum_{b\in \mathcal{S}} p_b \exp\big\{-\gamma \sum_{i=1}^N|z_{i,k} - h_{i,k} b|^2 \big\}$. The variational mean and variance of the posterior distribution in \eqref{postDist} can be calculated as
\begin{align}
    \ms{F}_s(\mb{z}_k,\mb{h}_k,\gamma) &= \sum_{a\in \mathcal{S}} a \, p(s_k=a|\mb{z}_k;\mb{h}_k,\gamma),\\
    \ms{G}_s(\mb{z}_k,\mb{h}_k,\gamma) &= \sum_{a\in \mathcal{S}}  |a|^2 \, p(s_k=a|\mb{z}_k;\mb{h}_k,\gamma) \nonumber \\
    &\quad - |{\sf F}_s(\mb{z}_k,\mb{h}_k,\gamma)|^2.
\end{align}

\def\baselinestretch{.99}
\bibliographystyle{IEEEtran}

\begin{thebibliography}{10}
	\providecommand{\url}[1]{#1}
	\csname url@samestyle\endcsname
	\providecommand{\newblock}{\relax}
	\providecommand{\bibinfo}[2]{#2}
	\providecommand{\BIBentrySTDinterwordspacing}{\spaceskip=0pt\relax}
	\providecommand{\BIBentryALTinterwordstretchfactor}{4}
	\providecommand{\BIBentryALTinterwordspacing}{\spaceskip=\fontdimen2\font plus
		\BIBentryALTinterwordstretchfactor\fontdimen3\font minus
		\fontdimen4\font\relax}
	\providecommand{\BIBforeignlanguage}[2]{{%
			\expandafter\ifx\csname l@#1\endcsname\relax
			\typeout{** WARNING: IEEEtran.bst: No hyphenation pattern has been}%
			\typeout{** loaded for the language `#1'. Using the pattern for}%
			\typeout{** the default language instead.}%
			\else
			\language=\csname l@#1\endcsname
			\fi
			#2}}
	\providecommand{\BIBdecl}{\relax}
	\BIBdecl
	
	\bibitem{van2024Asilomar}
	T.-V. Nguyen, S.~Nassirpour, I.~Atzeni, A.~T{\"o}lli, A.~L. Swindlehurst, and
	D.~H.~N. Nguyen, ``Optimal detection in {MIMO} systems using spatial
	sigma-delta {ADCs},'' in \emph{Proc. Asilomar Conf. Signals, Syst., and
		Comput. (ASILOMAR)}, 2024.
	
	\bibitem{rajatheva2020white}
	N.~Rajatheva, I.~Atzeni, E.~Bjornson, A.~Bourdoux, S.~Buzzi, J.-B. Dore,
	S.~Erkucuk, M.~Fuentes, K.~Guan \emph{et~al.}, ``White paper on broadband
	connectivity in {6G},'' \emph{arXiv preprint arXiv:2004.14247}, 2020.
	
	\bibitem{nguyen2017hybrid}
	D.~H.~N. Nguyen, L.~B. Le, T.~Le-Ngoc, and R.~W. Heath, ``Hybrid {MMSE}
	precoding and combining designs for mmwave multiuser systems,'' \emph{IEEE
		Access}, vol.~5, pp. 19\,167--19\,181, 2017.
	
	\bibitem{buzzi2018energy}
	S.~Buzzi and C.~D’Andrea, ``Energy efficiency and asymptotic performance
	evaluation of beamforming structures in doubly massive {MIMO} {mmWave}
	systems,'' \emph{IEEE Trans. Green Commun. Netw}, vol.~2, no.~2, pp.
	385--396, 2018.
	
	\bibitem{sohrabi2018one}
	F.~Sohrabi, Y.-F. Liu, and W.~Yu, ``One-bit precoding and constellation range
	design for massive {MIMO} with {QAM} signaling,'' \emph{IEEE J. Sel. Top.
		Signal Process.}, vol.~12, no.~3, pp. 557--570, 2018.
	
	\bibitem{jedda2018}
	H.~Jedda, A.~Mezghani, A.~L. Swindlehurst, and J.~A. Nossek, ``Quantized
	constant envelope precoding with {PSK and QAM} signaling,'' \emph{IEEE Trans.
		Wireless Commun.}, vol.~17, no.~12, pp. 8022--8034, 2018.
	
	\bibitem{fesl2023mean}
	B.~Fesl, M.~Koller, and W.~Utschick, ``On the mean square error optimal
	estimator in one-bit quantized systems,'' \emph{IEEE Trans. Signal Process.},
	vol.~71, pp. 1968--1980, 2023.
	
	\bibitem{roth2018comparison}
	K.~Roth, H.~Pirzadeh, A.~L. Swindlehurst, and J.~A. Nossek, ``A comparison of
	hybrid beamforming and digital beamforming with low-resolution {ADCs} for
	multiple users and imperfect {CSI},'' \emph{IEEE J. Sel. Top. Signal
		Process.}, vol.~12, no.~3, pp. 484--498, 2018.
	
	\bibitem{safa2024data}
	K.~Safa, R.~Combes, R.~De~Lacerda, and S.~Yang, ``Data detection in 1-bit
	quantized {MIMO} systems,'' \emph{IEEE Trans. Commun.}, vol.~72, no.~9, pp.
	5396--5410, Sep. 2024.
	
	\bibitem{nguyen2023variational}
	L.~V. Nguyen, A.~L. Swindlehurst, and D.~H.~N. Nguyen, ``Variational {B}ayes
	for joint channel estimation and data detection in few-bit massive {MIMO}
	systems,'' \emph{IEEE Trans. Signal Process.}, pp. 3408--3423, July 2024.
	
	\bibitem{li2017channel}
	Y.~Li, C.~Tao, G.~Seco-Granados, A.~Mezghani, A.~L. Swindlehurst, and L.~Liu,
	``Channel estimation and performance analysis of one-bit massive {MIMO}
	systems,'' \emph{IEEE Trans. Signal Process.}, vol.~65, no.~15, pp.
	4075--4089, 2017.
	
	\bibitem{zhang2016spectral}
	J.~Zhang, L.~Dai, S.~Sun, and Z.~Wang, ``On the spectral efficiency of massive
	{MIMO} systems with low-resolution {ADCs},'' \emph{IEEE Commun. Lett.},
	vol.~20, no.~5, pp. 842--845, 2016.
	
	\bibitem{fan2015uplink}
	L.~Fan, S.~Jin, C.-K. Wen, and H.~Zhang, ``Uplink achievable rate for massive
	{MIMO} systems with low-resolution {ADC},'' \emph{IEEE Commun. Lett.},
	vol.~19, no.~12, pp. 2186--2189, 2015.
	
	\bibitem{saxena2017analysis}
	A.~K. Saxena, I.~Fijalkow, and A.~L. Swindlehurst, ``Analysis of one-bit
	quantized precoding for the multiuser massive {MIMO} downlink,'' \emph{IEEE
		Trans. Signal Process.}, vol.~65, no.~17, pp. 4624--4634, 2017.
	
	\bibitem{bazrafkan2020asymptotic}
	A.~Bazrafkan and N.~Zlatanov, ``Asymptotic capacity of massive {MIMO} with
	1-bit {ADCs} and 1-bit {DACs} at the receiver and at the transmitter,''
	\emph{IEEE Access}, vol.~8, pp. 152\,837--152\,850, 2020.
	
	\bibitem{nguyen2021svm}
	L.~V. Nguyen, A.~L. Swindlehurst, and D.~H.~N. Nguyen, ``{SVM}-based channel
	estimation and data detection for one-bit massive {MIMO} systems,''
	\emph{IEEE Trans. Signal Process.}, vol.~69, pp. 2086--2099, 2021.
	
	\bibitem{de2010sigma}
	J.~M. de~la Rosa, ``Sigma-delta modulators: {T}utorial overview, design guide,
	and state-of-the-art survey,'' \emph{IEEE Trans. Circuits Syst. I: Regul.
		Pap.}, vol.~58, no.~1, pp. 1--21, 2010.
	
	\bibitem{rao2021massive}
	S.~Rao, G.~Seco-Granados, H.~Pirzadeh, J.~A. Nossek, and A.~L. Swindlehurst,
	``Massive {MIMO} channel estimation with low-resolution spatial sigma-delta
	{ADCs},'' \emph{IEEE Access}, vol.~9, pp. 109\,320--109\,334, 2021.
	
	\bibitem{aziz1996overview}
	P.~M. Aziz, H.~V. Sorensen, and J.~Van~der Spiegel, ``An overview of
	sigma-delta converters,'' \emph{IEEE Signal Process. Mag.}, vol.~13, no.~1,
	pp. 61--84, 1996.
	
	\bibitem{pirzadeh2020spectral}
	H.~Pirzadeh, G.~Seco-Granados, S.~Rao, and A.~L. Swindlehurst, ``Spectral
	efficiency of one-bit sigma-delta massive {MIMO},'' \emph{IEEE J. Sel. Areas
		Commun.}, vol.~38, no.~9, pp. 2215--2226, 2020.
	
	\bibitem{sankar2022channel}
	R.~P. Sankar and S.~P. Chepuri, ``Channel estimation in {MIMO} systems with
	one-bit spatial sigma-delta {ADCs},'' \emph{IEEE Trans. Signal Process.},
	vol.~70, pp. 4681--4696, 2022.
	
	\bibitem{atzeni2023doubly}
	I.~Atzeni, A.~T{\"o}lli, D.~H.~N. Nguyen, and A.~L. Swindlehurst, ``Doubly
	1-bit quantized massive {MIMO},'' in \emph{Proc. Asilomar Conf. Signals,
		Syst., and Comput. (ASILOMAR)}, 2023.
	
	\bibitem{jeon2018one}
	Y.-S. Jeon, N.~Lee, S.-N. Hong, and R.~W. Heath, ``One-bit sphere decoding for
	uplink massive {MIMO} systems with one-bit {ADCs},'' \emph{IEEE Trans.
		Wireless Commun.}, vol.~17, no.~7, pp. 4509--4521, 2018.
	
	\bibitem{jeon2019robust}
	Y.-S. Jeon, N.~Lee, and H.~V. Poor, ``Robust data detection for {MIMO} systems
	with one-bit {ADCs}: {A} reinforcement learning approach,'' \emph{IEEE Trans.
		Wireless Commun.}, vol.~19, no.~3, pp. 1663--1676, 2019.
	
	\bibitem{nguyen2020svm}
	L.~V. Nguyen, D.~H. Nguyen, and A.~L. Swindlehurst, ``{SVM}-based channel
	estimation and data detection for massive {MIMO} systems with one-bit
	{ADCs},'' in \emph{Proc. IEEE Int. Conf. Commun. (ICC)}, 2020.
	
	\bibitem{jeon2018supervised}
	Y.-S. Jeon, S.-N. Hong, and N.~Lee, ``Supervised-learning-aided communication
	framework for {MIMO} systems with low-resolution {ADCs},'' \emph{IEEE Trans.
		Veh. Technol.}, vol.~67, no.~8, pp. 7299--7313, 2018.
	
	\bibitem{khobahi2021lord}
	S.~Khobahi, N.~Shlezinger, M.~Soltanalian, and Y.~C. Eldar, ``{LoRD-Net}:
	Unfolded deep detection network with low-resolution receivers,'' \emph{IEEE
		Trans. Signal Process.}, vol.~69, pp. 5651--5664, 2021.
	
	\bibitem{azizzadeh2019ber}
	A.~Azizzadeh, R.~Mohammadkhani, S.~V. A.-D. Makki, and E.~Bj{\"o}rnson, ``{BER}
	performance analysis of coarsely quantized uplink massive {MIMO},''
	\emph{Signal Process.}, vol. 161, pp. 259--267, Aug. 2019.
	
	\bibitem{liu2023achievable}
	L.~Liu, Y.~Chi, Y.~Li, and Z.~Zhang, ``Achievable rates of generalized linear
	systems with orthogonal/vector amp receiver,'' \emph{IEEE Trans. Signal
		Process.}, vol.~71, pp. 4116--4133, Nov. 2023.
	
	\bibitem{tian2022generalized}
	F.~Tian, L.~Liu, and X.~Chen, ``Generalized memory approximate message passing
	for generalized linear model,'' \emph{IEEE Trans. Signal Process.}, vol.~70,
	pp. 6404--6418, Oct. 2022.
	
	\bibitem{fernandes2022multiuser}
	A.~B.~L. Fernandes, Z.~Shao, L.~T. Landau, and R.~C. de~Lamare,
	``Multiuser-mimo systems using comparator network-aided receivers with 1-bit
	quantization,'' \emph{IEEE Trans. on Commun.}, vol.~71, no.~2, pp. 908--922,
	2022.
	
	\bibitem{chen2024low}
	Z.~Chen, W.-X. Long, R.~Chen, and M.~Moretti, ``Low-cost {OAM} spatial
	oversampling receiver with 1-bit quantized comparators and {DL}-based
	detector,'' \emph{IEEE Commun. Lett.}, vol.~28, no.~6, Mar. 2024.
	
	\bibitem{nguyen2022variational}
	\BIBentryALTinterwordspacing
	D.~H.~N. Nguyen, I.~Atzeni, A.~T{\"o}lli, and A.~L. Swindlehurst, ``A
	variational {B}ayesian perspective on massive {MIMO} detection,'' \emph{arXiv
		preprint arXiv:2205.11649}, 2022. [Online]. Available:
	\url{https://arxiv.org/pdf/2205.11649}
	\BIBentrySTDinterwordspacing
	
	\bibitem{zhu2019grid}
	J.~Zhu, C.-k. Wen, J.~Tong, C.~Xu, and S.~Jin, ``Grid-less variational
	{B}ayesian channel estimation for antenna array systems with low resolution
	{ADCs},'' \emph{IEEE Trans. Wireless Commun.}, vol.~19, no.~3, pp.
	1549--1562, Mar. 2019.
	
	\bibitem{palguna2016millimeter}
	D.~S. Palguna, D.~J. Love, T.~A. Thomas, and A.~Ghosh, ``Millimeter wave
	receiver design using low precision quantization and parallel
	{$\Delta\Sigma$} architecture,'' \emph{IEEE Trans. Wireless Commun.},
	vol.~15, no.~10, pp. 6556--6569, 2016.
	
	\bibitem{shao2019one}
	M.~Shao, W.-K. Ma, Q.~Li, and A.~L. Swindlehurst, ``One-bit sigma-delta {MIMO}
	precoding,'' \emph{IEEE J. Sel. Top. Signal Process.}, vol.~13, no.~5, pp.
	1046--1061, 2019.
	
	\bibitem{corey2016spatial}
	R.~M. Corey and A.~C. Singer, ``Spatial sigma-delta signal acquisition for
	wideband beamforming arrays,'' in \emph{Proc. Int. ITG Workshop Smart
		Antennas (WSA)}, 2016.
	
	\bibitem{keung2024transmitting}
	W.-Y. Keung, H.~V. Cheng, and W.-K. Ma, ``Transmitting data through
	reconfigurable intelligent surface: {A} spatial sigma-delta modulation
	approach,'' in \emph{Proc. IEEE Int. Conf. Acoust., Speech, and Signal
		Process. (ICASSP)}, 2024.
	
	\bibitem{krieger2013dense}
	J.~D. Krieger, C.-P. Yeang, and G.~W. Wornell, ``Dense delta-sigma phased
	arrays,'' \emph{IEEE Trans. Antennas Propag.}, vol.~61, no.~4, pp.
	1825--1837, 2013.
	
	\bibitem{andrews2016modeling}
	J.~G. Andrews, T.~Bai, M.~N. Kulkarni, A.~Alkhateeb, A.~K. Gupta, and R.~W.
	Heath, ``Modeling and analyzing millimeter wave cellular systems,''
	\emph{IEEE Trans. Commun.}, vol.~65, no.~1, pp. 403--430, 2016.
	
	\bibitem{demir2020bussgang}
	O.~T. Demir and E.~Bjornson, ``The bussgang decomposition of nonlinear systems:
	Basic theory and mimo extensions [lecture notes],'' \emph{IEEE Signal
		Process. Mag.}, vol.~38, no.~1, pp. 131--136, 2020.
	
	\bibitem{alkhateeb2014channel}
	A.~Alkhateeb, O.~El~Ayach, G.~Leus, and R.~W. Heath, ``Channel estimation and
	hybrid precoding for millimeter wave cellular systems,'' \emph{IEEE J. Sel.
		Top. Signal Process.}, vol.~8, no.~5, pp. 831--846, 2014.
	
	\bibitem{jin2020one}
	B.~Jin, J.~Zhu, Q.~Wu, Y.~Zhang, and Z.~Xu, ``One-bit {LFMCW} radar: Spectrum
	analysis and target detection,'' \emph{IEEE Trans. Aerosp. Electron. Syst.},
	vol.~56, no.~4, pp. 2732--2750, 2020.
	
	\bibitem{thoota2021variational}
	S.~S. Thoota and C.~R. Murthy, ``Variational {Bayes} joint channel estimation
	and soft symbol decoding for uplink massive {MIMO} systems with low
	resolution {ADCs},'' \emph{IEEE Trans. Commun.}, vol.~69, no.~5, pp.
	3467--3481, 2021.
	
	\bibitem{nguyen2022deep}
	L.~V. Nguyen, D.~H. Nguyen, and A.~L. Swindlehurst, ``Deep learning for
	estimation and pilot signal design in few-bit massive {MIMO} systems,''
	\emph{IEEE Trans. Wireless Commun.}, vol.~22, no.~1, pp. 379--392, 2022.
	
	\bibitem{heath2016overview}
	R.~W. Heath, N.~Gonzalez-Prelcic, S.~Rangan, W.~Roh, and A.~M. Sayeed, ``An
	overview of signal processing techniques for millimeter wave {MIMO}
	systems,'' vol.~10, no.~3, pp. 436--453, Apr. 2016.
	
	\bibitem{kang2023deep}
	J.-M. Kang, ``Deep learning-based robust channel estimation for {MIMO IoT}
	systems,'' \emph{IEEE Internet Things J.}, vol.~11, no.~6, pp. 9882 -- 9895,
	Mar. 2024.
	
	\bibitem{zhang2022mmwave}
	H.~Zhang, Y.~Zhang, J.~Cosmas, N.~Jawad, W.~Li, R.~Muller, and T.~Jiang,
	``mmwave indoor channel measurement campaign for 5g new radio indoor
	broadcasting,'' \emph{IEEE Trans. Broadcast.}, vol.~68, no.~2, pp. 331--344,
	Jun. 2022.
	
	\bibitem{hussain2022integrated}
	N.~Hussain and N.~Kim, ``Integrated microwave and mm-wave mimo antenna module
	with 360 pattern diversity for 5g internet of things,'' \emph{IEEE Internet
		Things J.}, vol.~9, no.~24, pp. 24\,777--24\,789, Dec. 2022.
	
	\bibitem{pirzadehmc}
	H.~Pirzadeh, G.~Seco-Granados, A.~L. Swindlehurst, and J.~A. Nossek, ``On the
	effect of mutual coupling in one-bit spatial sigma-delta massive {MIMO}
	systems,'' in \emph{Proc. IEEE Int. Workshop Signal Process. Adv. in Wireless
		Commun. (SPAWC)}, 2020.
	
	\bibitem{bussgang1952crosscorrelation}
	J.~J. Bussgang, ``Crosscorrelation functions of amplitude-distorted gaussian
	signals,'' vol. 216, 1952.
	
	\bibitem{pirzadeh2020effect}
	H.~Pirzadeh, G.~Seco-Granados, A.~L. Swindlehurst, and J.~A. Nossek, ``On the
	effect of mutual coupling in one-bit spatial {Sigma-Delta} massive {MIMO}
	systems,'' in \emph{Proc. IEEE Int. Workshop Signal Process. Adv. in Wireless
		Commun. (SPAWC)}, 2020.
	
	\bibitem{bishop2006pattern}
	C.~M. Bishop and N.~M. Nasrabadi, \emph{Pattern Recognition and Machine
		Learning}.\hskip 1em plus 0.5em minus 0.4em\relax Springer, 2006, vol.~4,
	no.~4.
	
	\bibitem{wainwright2008graphical}
	M.~W. et~al, ``{Graphical Models, Exponential Families, and Variational
		Inference},'' \emph{Found. and Trends Mach. Learn.}, vol.~1, no. 1--2, pp.
	1--305, 2008.
	
\end{thebibliography}

\end{document}